\documentclass[prx,aps,amsart,amssymb,epsfig,showpacs,twocolumn,eqsecnum,superscriptaddress]{revtex4-1}

\usepackage{amsmath,amssymb,amsfonts,graphicx,multirow,color,bm,textcomp,mathtools}
\usepackage{paralist}
\usepackage{srcltx}
\usepackage[all,cmtip]{xy}
\usepackage{mathrsfs}
\usepackage{srcltx,soul,subfigure}
\usepackage{threeparttable,dsfont,bbm}

\usepackage{hyperref}
\usepackage{cleveref}

\usepackage{soul}

\crefname{equation}{Eq.}{Eqs.}

\newcommand{\ket}[1]{| #1 \rangle}

\newcommand{\bee}{\begin{eqnarray}}
\newcommand{\ee}{\end{eqnarray}}
\newcommand{\bma}{\begin{pmatrix}}
\newcommand{\ema}{\end{pmatrix}}
\newcommand{\balig}{\begin{align}}
\newcommand{\ealig}{\end{align}}

\newcommand{\ba}{\begin{align}}
\newcommand{\ea}{\end{align}}

\newcommand{\ignore}[1]{}

\newcommand{\lt}{\vec{t}} 
\newcommand{\kk}{\vec{k}} 
\newcommand{\Rb}{\vec{R}} 

\newcommand\Tstrut{\rule{0pt}{2.6ex}}         
\newcommand\Bstrut{\rule[-0.9ex]{0pt}{0pt}}   

\usepackage{dcolumn}
\newcolumntype{C}[1]{>{\centering\let\newline\\\arraybackslash\hspace{0pt}}m{#1}}

\begin{document}

\title{Symmetry-enforced band crossings in trigonal materials: \\ Accordion states and Weyl nodal lines}

\author{Y.-H. Chan}
\email{yanghao@umich.edu}
\affiliation{Institute of Atomic and Molecular Sciences, Academia Sinica, Taipei 10617, Taiwan}
\affiliation{Materials Sciences Division, Lawrence Berkeley National Laboratory, Berkeley, CA 94720, USA}

\author{Berkay Kilic}
\affiliation{Max-Planck-Institut f\"ur Festk\"orperforschung, Heisenbergstrasse 1, D-70569 Stuttgart, Germany} 

\author{Moritz~M.~Hirschmann}
\affiliation{Max-Planck-Institut f\"ur Festk\"orperforschung, Heisenbergstrasse 1, D-70569 Stuttgart, Germany} 
  
\author{Ching-Kai Chiu}
\affiliation{Kavli Institute for Theoretical Sciences, University of Chinese Academy of Sciences, Beijing 100190, China}

\author{Leslie~M.~Schoop}
\affiliation{Department of Chemistry, Princeton University, Princeton, NJ 08544, USA}

\author{Darshan G. Joshi}
\affiliation{Max-Planck-Institut f\"ur Festk\"orperforschung, Heisenbergstrasse 1, D-70569 Stuttgart, Germany}  

\author{Andreas P. Schnyder}
\email{a.schnyder@fkf.mpg.de}
\affiliation{Max-Planck-Institut f\"ur Festk\"orperforschung, Heisenbergstrasse 1, D-70569 Stuttgart, Germany}

\begin{abstract}
Nonsymmoprhic symmetries, such as screw rotations or glide reflections, can  enforce band crossings within high-symmetry lines or planes of the Brillouin zone. 
When these band degeneracies are close to the Fermi energy, they can give rise to a number of unusual phenomena: 
e.g., anomalous magnetoelectric responses, transverse Hall currents, and exotic surface states. 
In this paper, we present a comprehensive classification of such nonsymmorphic band crossings in trigonal materials with 
strong spin-orbit coupling. We find that in trigonal systems there are two different types of nonsymmorphic band
degeneracies: (i) Weyl points protected by screw rotations with an accordion-like dispersion, 
and (ii) Weyl nodal lines protected by glide reflections.
We report a number of existing materials, where these band crossings are realized near the Fermi energy.
This includes Cu$_2$SrSnS$_4$ and elemental tellurium (Te), which
exhibit accordion Weyl points; and the tellurium-silicon clathrate  Te$_{16}$Si$_{38}$, which shows Weyl nodal lines.
The \emph{ab-initio} band structures and surface states of these materials are studied in detail,
and implications for experiments are briefly discussed.
\end{abstract}

\date{\rm\today}

\maketitle


\section{Introduction}

The famous non-crossing theorem by Wigner and von Neumann~\cite{neumann_wigner}
states that Bloch states with the same symmetry cannot be degenerate at a generic point in the Brillouin zone (BZ),
which prevents the formation of band crossings.
Instead, when two Bloch bands of the same symmetry approach each other at 
a generic momentum, they start to hybridize and undergo an avoided level crossing.
However, the non-crossing theorem does not apply to 
bands with non-trivial wavefunction topology, which can form topologically protected
band degeneracies~\cite{chiu_RMP_16,volovikLectNotes13,armitage_mele_vishwanath_review,burkov_review_Weyl,yang_ali_review_ndoal_line}.
When these band crossings are in the vicinity of the Fermi energy, 
they lead to a range of interesting phenomena, such as arc and drumhead surface states~\cite{WanVishwanathSavrasovPRB11,BurkovBalentsPRB11,xu_hasan_Na3Bi_fermi_arc_science_15,huang_hasan_TaAs_Fermi_Arc_nat_comm_15,nodal_line_Yang,bian_hasan_PbTaSe2_drumhead_nat_comm_16,bian_hasan_TlTaSe2_drumhead_PRB_16},
transverse topological currents~\cite{rui_schnyder_PRB_18,armitage_mele_vishwanath_review,burkov_review_Weyl}, and anomalous magnetoelectric responses~\cite{bzdusek_soluyanov_nature_16},
which could potentially be utilized for device applications~\cite{shi_wu_APL_15,valleytronics_review}. 

There are two different types of topological band crossings, namely,
accidental band crossings and symmetry-enforced band crossings.
The former are protected by symmorphic crystal symmetries 
and are only perturbatively stable~\cite{chiu_schnyder_PRB_14,shiozaki_sato_classification_PRB_14,zhao_PT_PRL_16}. That is, they can be adiabatically removed
by large symmetry-preserving deformations of the Hamiltonian, for example, through pair annihilation. 
Examples of accidental band crossings include
Dirac points and Dirac lines that are protected by space-time inversion, reflection, 
or rotation symmetry~\cite{castro_neto_geim_review_graphene_RMP_09,dirac_sm_Na3Bi_fang_zhong_PRB_12,chiu_schnyder_Na3Bi_2015,Dirac_SM_Cd3As2_fang_zhong_PRB_13,xie_schoop_Ca3P2_apl_15,nodal_line_Yamakage}.
Symmetry-enforced band crossings~\cite{young_kane_rappe_Dirac_3D_PRL_12,schoop_zrsis,bzdusek_soluyanov_nature_16,zhao_schnyder_PRB_16,michel_zak_PRB_99,young_kane_non_symmorphic_PRL_15,alexandradinata_hourglass_surface_PRX_16,furusaki_non_symmorphic_17,ryo_murakami_PRB_17,yang_furusaki_PRB_2017,fang_kee_fu_off_center_PRB_15,malard_johannesson_arXiv_18,zhang_hexagonals_PRMAT_18}, on the other hand, 
arise in the presence of \emph{nonsymmorphic} symmetries 
and are globally stable, i.e., they cannot be removed even
by large deformations of the Hamiltonian.
In other words, these band crossings are required to exist 
 due to nonsymmorphic symmetries alone, independent of 
the chemical composition and other details of the material.  
This fact allows us to construct the following strategy 
to discover new topological semimetals,
which consists of three steps: 
(i) identify the space groups (SGs) whose nonsymmorphic symmetries 
enforce  the desired band crossings,  
(ii) perform a database search for materials in these SGs,
and (iii) compute the electronic band structure of these materials to check whether 
the band crossings are near the Fermi energy. 
Previously, we applied this strategy to discover new topological semimetals with hexagonal symmetries~\cite{zhang_hexagonals_PRMAT_18}.
Here, we extend this analysis to trigonal systems.

\begin{table*}[t!]
\centering
\begin{ruledtabular}
\begin{tabular}{ l c c c c c c}  
\text{Space group} 
& 
Weyl nodal points
& 
Weyl nodal lines
& 
 Filling constraint & Materials \\ \hline
144 \ $(P3_1)$  & $\Gamma$$\Delta$A (6)  & -- & 6$\mathbb{N}$ &   Cu$_2$SrSnS$_4$   
\\ \hline
145 \ $(P3_2)$  & $\Gamma$$\Delta$A (6) & --  & 6$\mathbb{N}$ & Cu$_2$SrGeS$_4$
\\ \hline
151 \ $(P3_112)$  & $\Gamma$$\Delta$A (6) & -- & 6$\mathbb{N}$ & Ag$_2$HPO$_4$ 
\\ \hline
152 \ $(P3_121)$  & $\Gamma$$\Delta$A (6) & -- & 6$\mathbb{N}$ & Te, Se
\\ \hline
153 \ $(P3_212)$  & $\Gamma$$\Delta$A (6) & -- & 6$\mathbb{N}$ & -- \\ \hline
154  \ $(P3_221)$   & $\Gamma$$\Delta$A (6) & -- & 6$\mathbb{N}$ & --
\\ \hline
158 \ $(P3c1)$  & -- &  $k_x k_z$-plane   & 4$\mathbb{N}$ &  --   \\ \hline
159 \ $(P31c)$ & -- &  $k_x=\pi$    & 4$\mathbb{N}$ &  --  
\\ \hline 
161 \ $(R3c)$  & -- &  $k_{x}=-k_{y}$     & 4$\mathbb{N}$ &  Te$_{16}$Si$_{38}$
\\  
\end{tabular}
\end{ruledtabular}
\caption{
Classification of nonsymmorphic band crossings in trigonal materials with time-reversal symmetry and strong spin-orbit coupling.
The first column lists the trigonal space groups that exhibit topological band crossings.
The second and third columns indicate the high-symmetry lines and planes in which Weyl nodal points and Weyl nodal lines appear, respectively. 
The definition of the coordinate system and high-symmetry labels in the Brillouin zones are given in Fig.~\ref{mFig1}(a) and Fig.~\ref{mFig2}.
The electron fillings  $\nu$ for which a band insulator is possible,   are given in the fourth column, where
$m \mathbb{N}$ denotes the set $\{ m, 2m, 3m, \cdots \}$.
Example materials which realize the predicted band crossings are listed in the last column. 
\label{mTab1}
}
\end{table*}
 
There has been a number of previous works, which considered specific types of nonsymmorphic band crossings~\cite{bzdusek_soluyanov_nature_16,furusaki_non_symmorphic_17,ryo_murakami_PRB_17,yang_furusaki_PRB_2017} and proposed
  candidate materials, where these are realized~\cite{chen_vishwanath_nat_phys_17,bradlyn_bernevig_nature_17,schoop_CeSbTe_2018,topp_schoop_ZrSiTe_16}. However, 
a comprehensive classification of all possible nonsymmorphic band-crossings in trigonal materials has not been performed before. 
We find that among the 25 trigonal SGs, six support Weyl points with accordion-like dispersions
and three support Weyl nodal lines (see Table~\ref{mTab1}). We identify several materials, 
 where these topological band crossings are realized. This includes
 elemental tellurium (Te) with accordion Weyl points,
 and the trigonal high-temperature form of Te$_{16}$Si$_{38}$ with Weyl nodal lines.
As a byproduct of our analysis, we also obtain the tight  filling constraints for the existence of band insulators, 
which are in agreement with the recent literature~\cite{parameswaran_nature_2013,watanabe_vishwanath_PRL_16}.
Furthermore, we find that in trigonal materials, as opposed to hexagonal materials~\cite{zhang_hexagonals_PRMAT_18}, there exist no band crossings protected by off-centered symmetries~\cite{yang_furusaki_PRB_2017,fang_kee_fu_off_center_PRB_15,malard_johannesson_arXiv_18}.
We note that our analysis takes into account the full connectivity of the bands in the entire BZ, which goes
beyond the results obtained through symmetry indicators~\cite{materiae_zhang_fang_Nature_19,vergniory_catalogue_Nature_19,tang_vishwanath_catalogue_Nature_19}.

The remainder of this paper is organized as follows. In Sec.~\ref{mSec2} we 
classify Weyl points protected by screw rotation symmetries
and discuss how symmetries constrain the multiplicities
of the Weyl points. 
Section~\ref{mSec3} is devoted
to the study of Weyl nodal lines protected by glide reflection.
In Sec.~\ref{sec_filling_constraints} we discuss
the filling constraints imposed by the nonsymmorphic symmetries. 
In Sec.~\ref{mSec4}  we present the example materials
that realize the predicted band crossings.
Our summary and conclusions   are given in Sec.~\ref{sec_conclusion}. 
In Appendix~\ref{appencix_band_structure} we present additional figures of the band structure calculations.
The calculated surface states of tellurium are shown in Appendix~\ref{appendixB}.

\section{Non-symmorphic nodal points}
\label{mSec2}

In this section we show how three-fold screw-rotation symmetries in trigonal space groups lead to protected Weyl points.
To derive this we employ two different approaches:
(i) Using symmetry analysis we   show how the evolution of eigenvalues of the symmetry operator along a line in 
the BZ enforces band crossings (Sec.~\ref{sec_IIA}). (ii)~Using crystallographic group theory we   show how the
compatibility relations between irreducible representations (irreps) at different high-symmetry points of the BZ~\cite{michel_zak_phys_reports_01,kruthoff_slager_PRX_17,elcoro_aroyo_JAC_17} 
lead to enforced band crossings (Sec.~\ref{sec_IIB}).  
We also give a brief discussion on how time-reversal and SG symmetries
lead to higher multiplicities of the Weyl points (Sec.~\ref{secIIC}). 

\subsection{Symmetry eigenvalues}
\label{sec_IIA}

\begin{figure*}[th]
\centering
\includegraphics[width=.9\textwidth]{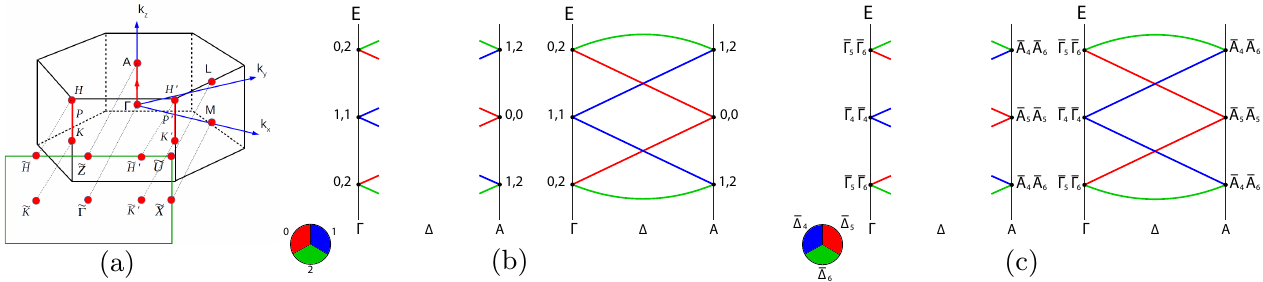}
\caption{
(a) Bulk BZ (black lines) for trigonal space groups with hexagonal lattice systems (P-trigonal).
The high-symmetry points and lines are indicated in red.
The green lines show the surface BZ for surfaces perpendicular to the (-110) direction. 
(b), (c) Kramers pairings and band connectivity diagrams for the $\Gamma - \Delta - A$ line of the hexagonal BZ,
which is left invariant by the three-fold screw rotation $C_{3,p}$.
Here, we focus on the case $p=1$, e.g., on SG No.~144 (P3$_1$). (Similar band crossings are also realized for $p=2$.)
The left panels in (b) and (c) indicate how the Bloch bands  $\ket{\psi_{m} (\kk)}$
pair up into Kramers partners at the time-reversal invariant momenta $\Gamma$ and $A$, see Tables~\ref{mTab2} and~\ref{mTab3}. 
The right panels in (b) and (c) show the band connectivity diagrams  along
the path $\Gamma - \Delta - A$.
Six bands form a connected group with at least two symmetry-enforced Weyl points, as explained in the text.
In (b) the bands are marked by their $C_{3,p}$ eigenvalue label $m$, while in
(c) they are marked by their double-valued irreps $\bar{\Delta}_i$. 
 }
\label{mFig1}
\label{fig:mWeyl} 
\end{figure*}

In general, a symmetry operator 
$G=\lbrace g|\lt \rbrace$ of a given space group is a combination of a point group symmetry $g$ and a lattice translation $\lt$. 
In case of a nonsymmorphic symmetry the lattice translation $\lt$ is some fraction of the Bravais-lattice vector. Here, we are interested in the 
three-fold screw rotation, which has the following form:
\begin{equation}
\label{eq:c3}
C_{3,p} : (x, y, z) \rightarrow (-y, x-y, z+\frac{p}{3}) \otimes (\frac{1}{2} \sigma_{0} - \frac{i\sqrt{3}}{2} \sigma_{z}) \,,
\end{equation}
such that $p \in  \lbrace 1, 2 \rbrace$.
Within the $g$-invariant space of the BZ, i.e. where $g \kk = \kk$, the Bloch states $\ket{\psi_{m} (\kk)} $ can be constructed 
to be simultaneous eigenfunctions of G and the Hamiltonian, satisfying the following eigenvalue equation:
\begin{equation}
\label{eq:c3_eigen}
C_{3,p} \ket{\psi_{m} (\kk)} = e^{i\pi(2m+1)/3} e^{-ipk_{z}/3} \ket{\psi_{m} (\kk)} \,,
\end{equation}
with $m \in \lbrace 0, 1, 2 \rbrace$.
The above equation follows from the fact that $C_{3,p}^{3} = - p \Rb$, where $\Rb$ is a Bravais lattice vector (here $\Rb=\hat{z}$) 
and the minus sign arises due to $2\pi$ rotation of the electron spin, relevant for spin-orbit coupling. 
From Eq.~\eqref{eq:c3_eigen}, we observe that the eigenvalues of $C_{3,p}$ evolve 
along the line segment $\kk = (0, 0, k_{z})$ with $k_{z} \in [0,\pi]$, i.e., along the path $\Gamma - \Delta - A$ in Fig. \ref{mFig1}, 
which is invariant under $C_{3}$ rotation. 

Note that the points $\Gamma$ and $A$ are time-reversal invariant momenta, where in the presence of time-reversal symmetry 
bands pair-up with their respective Kramers partner. In terms of the eigenvalues in Eq.~\eqref{eq:c3_eigen} this means 
that a band with real eigenvalue pairs with itself, while a band with complex eigenvalue pairs with the band
whose eigenvalue is the complex conjugate.
Using the label $m$ to identify the bands and corresponding eigenvalues, we find the following pairing scenario:
At the $\Gamma$ point ($k_{z} =0$) there are two Kramers pairs of bands, namely, $(1,1)$ and $(0,2)$, irrespective of the value of $p$. 
At the $A$ point ($k_{z} = \pi$), on the other hand, Kramers pairs depend on the value of $p$ and pair up as indicated in
Table~\ref{mTab2}. 
Since the bands pair up differently at $\Gamma$ and A, they
show, in the absence of additional symmetries, a nontrivial connectivity along the path $\Gamma - \Delta - A$, as displayed in Fig.~\ref{fig:mWeyl}(b).
We find that for both $p=1$ and $p=2$ there are six bands forming a connected group, which must cross at least two times,
leading to two Weyl point degeneracies with an accordion-like dispersion.

\begin{table}[t!]
\begin{ruledtabular}
\begin{tabular}{ c c }  
$p$ 
& 
Pairings of $m$ 
\\ \hline
1 & (0,0), (1,2) \\ \hline 
2 & (0,1), (2,2)
\end{tabular}
\end{ruledtabular}
\caption{
This table lists how the Bloch bands $\ket{\psi_{m} (\kk)} $ pair
up into Kramers partners at the A point of the BZ in materials 
with a three-fold screw rotation  $C_{3,p}$.
The pairing depends on $p$, i.e., the fractional lattice translation part of $C_{3,p}$, see Eq.~\eqref{eq:c3_eigen}.
The right column indicates the eigenvalue labels $m$ of the Kramers pairs
at the A point. 
\label{mTab2}
}
\end{table}

We note that the presence of additional symmetries such as inversion or mirror, which either commute or anti-commute with $C_{3,p}$, 
may lead to additional degeneracies at the high-symmetry points in the BZ. This can remove the above discussed Weyl points. 
In particular, out of the eighteen  trigonal space groups with hexagonal lattice system (P-trigonal) there are only six space groups without inversion symmetry
(SGs with Nos.\ 144, 145, 151, 152, 153, and 154).
These are the only SGs which host symmetry-enforced Weyl points with accordion-like dispersion, see Table~\ref{mTab1}.

In passing we note that 
the seven trigonal space groups with rhombohedral lattice system (R-trigonal)
do not have a nonsymmorphic three-fold screw-rotation symmetry, but just three-fold symmorphic
rotation symmetries (i.e. $p=0$ in $C_{3,p}$). Consequently, the eigenvalues of these symmorphic rotation symmetries are
momentum independent and hence there are no symmetry-enforced Weyl points for R-trigonal space groups.

\subsection{Compatibility relations between irreps}
\label{sec_IIB}

Next, we explain how the existence of symmetry-enforced
Weyl points can be derived form the compatibility relations of irreducible representations (irreps) at
different high-symmetry points/lines of the BZ~\cite{michel_zak_phys_reports_01,kruthoff_slager_PRX_17,elcoro_aroyo_JAC_17}. 
Here, we give a brief outline of this derivation for SG No.~144~(P3$_1$). Additional details can be found in Ref.~\cite{zhang_hexagonals_PRMAT_18}. 

Bloch bands $\ket{\psi_{m} (\kk)}$ restricted to a high-symmetry line (or point) can be labeled by the irreducible
representations (irreps) of the symmetry operators that leave this line (or point) invariant. 
An irrep of a given symmetry group can be viewed as a set of matrices which form a group that
is homomorphic to the symmetry group, 
with the property that there does not exist a unitary transformation that brings all these matrices into block diagonal form.
While these sets of matrices are not unique, their trace (i.e., their character) is unchanged under any unitary transformation. 
Therefore, for a given high-symmetry line (or point) in the BZ one can define a character table, which lists the irrpes
and their characters for all the symmetry operations that leave this line (or point) invariant.

Now, as we move from a high-symmetry point to a high-symmetry line in the BZ, the symmetry is lowered,
and hence, an irrep $A_{\sigma}$ of the high-symmetry point is, in general, no longer irreducible 
for the high-symmetry line. That is the irrep $A_{\sigma}$ of the high-symmetry point decomposes,
in general, into smaller irreps $B_{i,\sigma}$ of the high-symmetry line. 
However, under such a decomposition the character $\chi$ of each relevant symmetry   must be preserved, i.e.,  
\begin{equation}
\label{eq:chara}
\chi (A_{\sigma}) = \sum_{i=1}^{n} \chi (B_{i,\sigma}) \,.
\end{equation}
The above equation defines the {\em compatibility relations}
between the irreps $A_{\sigma}$ at a high-symmetry point  and
the irreps $B_{i,\sigma}$ at a high-symmetry line~\cite{bradley_irreps,miller_lover_irreps}.

\begin{table}[t!]
\begin{ruledtabular}
\begin{tabular}{ c c }  
Irreps/Sym. Element 
& 
${C}_{3}$
\\ \hline
$\bar{\Gamma}_{4}$ & $-1$ \Tstrut\\
$\bar{\Gamma}_{5}$ & $e^{-i\pi/3}$ \\
$\bar{\Gamma}_{6}$ & $e^{i\pi/3}$ \\ \hline 
$\bar{\Delta}_{4}$ & $e^{ik_{z}/3}$ \Tstrut\\
$\bar{\Delta}_{5}$ & $e^{i(k_{z} - \pi)/3}$ \\
$\bar{\Delta}_{6}$ & $e^{i(k_{z} + \pi)/3}$ \\ \hline 
$\bar{A}_{4}$ & $e^{-i2\pi/3}$ \Tstrut\\
$\bar{A}_{5}$ & $+1$ \\
$\bar{A}_{6}$ & $e^{i2\pi/3}$ 
\end{tabular}
\end{ruledtabular}
\caption{
Double-valued irreps for SG No.~144 (P3$_1$) without time-reversal symmetry
at the points $\Gamma$ and $A$ and at the line $\Delta$.
For the labelling of the irrpes we use the same convention
as in Ref.~\cite{elcoro_aroyo_JAC_17}.
\label{Tab144} \label{mTab3}
}
\end{table}

We shall now use these compatibility relations to study how the irreps decompose
in  SG No.~144 (P3$_1$) for the path $\Gamma - \Delta - A$, which is left invariant under the 
screw rotation ${C}_{3, 1}$.
Since we are interested in electronic band structures with spin-orbit coupling,
we need to consider the \emph{double-valued} irreps of SG No.~144 (P3$_1$) (see Table~\ref{Tab144}), as the identity 
operator is not equivalent to a rotation by $2\pi$ but rather by $4\pi$ on the Bloch sphere.
At the time-reversal invariant momenta (TRIMs), Bloch bands form Kramers pairs due to time-reversal symmetry. This means that at the TRIMs
pairs of complex conjugate irreps combine together to form a time-reversal symmetric irrep~\cite{bradley_irreps,miller_lover_irreps}. 
For the $1d$ irreps of  Table~\ref{Tab144} we can construct these time-reversal symmetric irreps by inspection.
At the $\Gamma$ point we find that the time-reversal symmetric irreps are given by the
direct product of $\bar{\Gamma}_{5}$ with 
$\bar{\Gamma}_{6}$  and $\bar{\Gamma}_{4}$ with itself. Similarly, at the $A$ point the time-reversal symmetric irreps 
are given by the direct product  of $\bar{A}_{4}$ with $\bar{A}_{6}$ and $\bar{A}_{5}$ with itself, see Fig.~\ref{mFig1}(c). 
As we move from the TRIMs $\Gamma$ and $A$ to a point on the high-symmetry line $\Delta$,
the two-dimensional time-reversal symmetric irreps decompose into one-dimensional irreps,
according to Eq.~\eqref{eq:chara}, see left part of Fig.~\ref{mFig1}(c). 
Connecting the respective irreps that split away from the points $\Gamma$ and $A$ we find
that the bands connect in a non-trivial way, with a minimum number of two
Weyl point crossings, as shown in the right part of Fig. \ref{mFig1}(c).
This is in exact agreement with the symmetry analysis presented above. 
Similar analysis can be done for the remaining five space groups of Table~\ref{mTab1} and we find perfect agreement therein.

\subsection{Weyl point multiplicities}
\label{secIIC}

\begin{figure}[t]
\begin{center}
\includegraphics[clip,width=0.9\columnwidth]{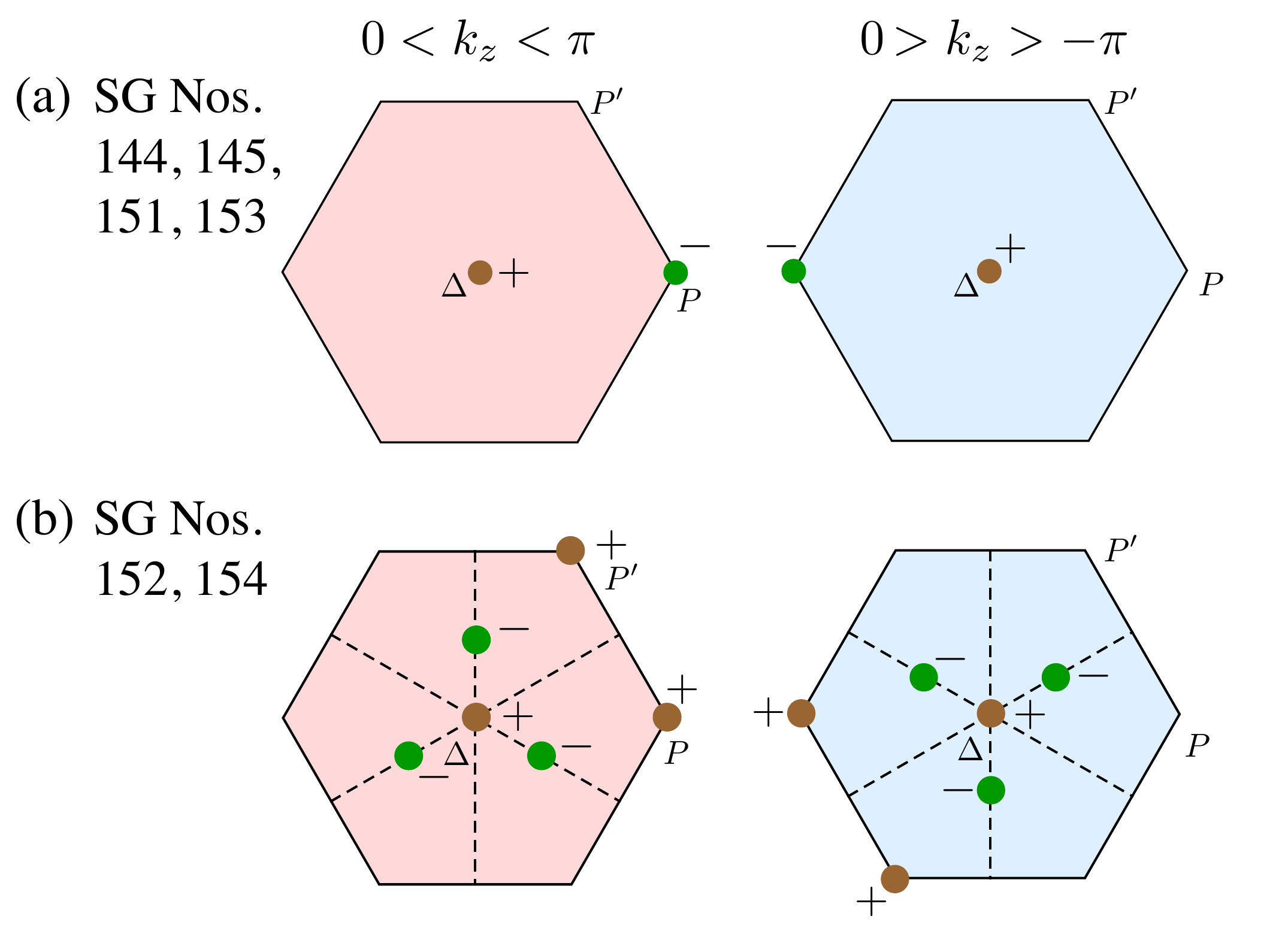}
\caption{
Weyl point configuration with minimal multiplicity for the six trigonal SGs with symmetry-enforced Weyl points (see Table~\ref{mTab1}). 
The red and blue hexagons represent the top view of the upper and lower parts of the hexagonal BZ,
which are related by time-reversal symmetry. The brown and green dots
indicate the positions of the Weyl points with positive and negative chirality, respectively.  
(a) In SG Nos.~144, 145, 151, and 153 the minimal Weyl point multiplicity is four. 
Weyl points with the same chirality are related by time-reversal symmetry.
(b)
In SG Nos.~152 and 154 the minimal Weyl point multiplicity is twelve. 
Weyl points with the same chirality are related by time-reversal symmetry and the mirror symmetries indicated
by the dashed lines.
 }
\label{Weyl_node_distribution} 
\end{center}
\end{figure}

So far we have only discussed the appearance of Weyl points on the $\Gamma - \Delta - A$ line.
However, it follows from symmetry arguments together with the fermion doubling theorem~\cite{NIELSEN198120}
that these cannot be the only Weyl points in the BZ.
This is because (i) there has to be another Weyl point away from the  $\Gamma - \Delta - A$ line with opposite
chirality, since the chiralities of all  Weyl points formed by one pair of bands must cancel out to zero~\cite{NIELSEN198120};
and (ii) time-reversal and SG symmetries lead to additional copies of the Weyl points that are located
away from the $\Gamma - \Delta - A$ line. 
Using these arguments, we now determine for each SG the minimum number
of Weyl points that is compatible with the symmetries and the fermion doubling theorem.
To do so, we first note that we can divide the BZ into two parts, 
an upper part with $k_z \in [0, \pi]$ and a lower part with
$k_z \in [-\pi, 0]$, which are related by time-reversal symmetry.
Hence, it is sufficient to analyze the Weyl point configuration in the
upper part; the configuration in the lower part can  be obtained by
applying time-reversal symmetry (see Fig.~\ref{Weyl_node_distribution}). 
For concreteness, we assume in the following
that the Weyl points at the $\Gamma - \Delta - A$ line have positive chirality $+1$.

\paragraph{SG Nos.~144, 145, 151, and 153.---}
Band structures in SG Nos.~144 and 145 possess only two
symmetries: the three-fold screw rotation along the $k_z$ axis $C_{3,p}$  and time-reversal
symmetry. In order to cancel the $+1$ chirality of the Weyl point at the $\Gamma - \Delta - A$ line,
we need to place a negative chirality Weyl point somewhere else in the BZ.
If we put it at an arbitrary position,  the  screw rotation $C_{3,p}$
would generate two more copies with the same chirality, leading to
an overcompensation. 
Instead, in order to construct a Weyl point configuration with minimal multiplicity, 
we need to put the negative chirality Weyl point at one of the other two $C_{3,p}$-invariant lines,
i.e., either on $K - P - H$ or on $K^\prime - P^\prime - H^\prime$.
With this, the chiralities  cancel out and we obtain a minimal Weyl point
multiplicity of four, see Fig.~\ref{Weyl_node_distribution}(a). 

SG Nos.~151 and 153 possess an additional two-fold screw rotation along the $k_y$ axis, which however
does not affect the above argument. Hence, also in these SGs the minimal Weyl point multiplicity is four.

\paragraph{SG Nos.~152 and 154.---}
SG Nos.~152 and 154 contain, besides the three-fold screw rotation $C_{3,p}$,  
a two-fold rotation $C_2$ along the (110) axis, which maps $(k_x,k_y,k_z) \to (k_x,-k_y,-k_z)$.
This $C_2$ rotation puts extra constraints on the multiplicity of the Weyl points. 
To find these constraints we combine the $C_2$ rotation with time-reversal symmetry, yielding the mirror symmetry 
$(k_x,k_y,k_z) \to (-k_x,k_y,k_z)$ [dashed lines in Fig.~\ref{Weyl_node_distribution}(b)]. 
If we now put a negative chirality Weyl point on the $K - P - H$ line, the mirror symmetry 
would generate another negative chirality Weyl point at the  $K^\prime - P^\prime - H^\prime$ line, leading to an 
overcancellation. 
Instead we place positive chirality Weyl points both at $K - P - H$  and $K^\prime - P^\prime - H^\prime$,
which we then compensate by negative chirality Weyl points on the three mirror planes, as shown in Fig.~\ref{Weyl_node_distribution}(b).
This configuration of twelve Weyl points has minimal multiplicity, 
satisfies all the symmetries, and has zero total chirality.

\begin{figure}[t!]
\centering
\subfigure[]{
\includegraphics[width = 0.432\columnwidth]{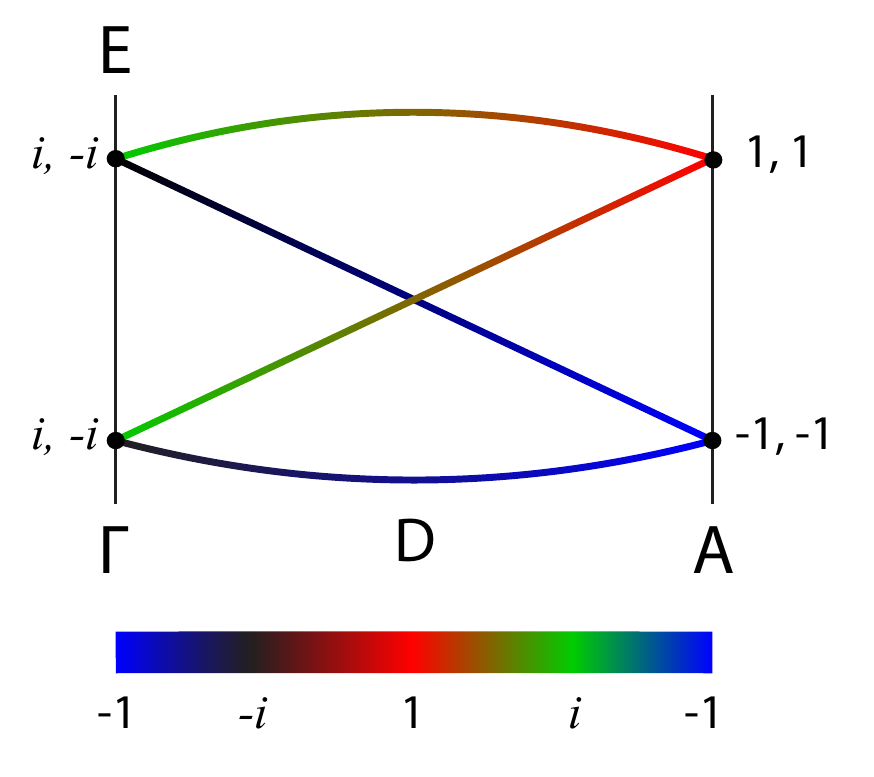}
}
\subfigure[]{
\includegraphics[width = 0.432\columnwidth]{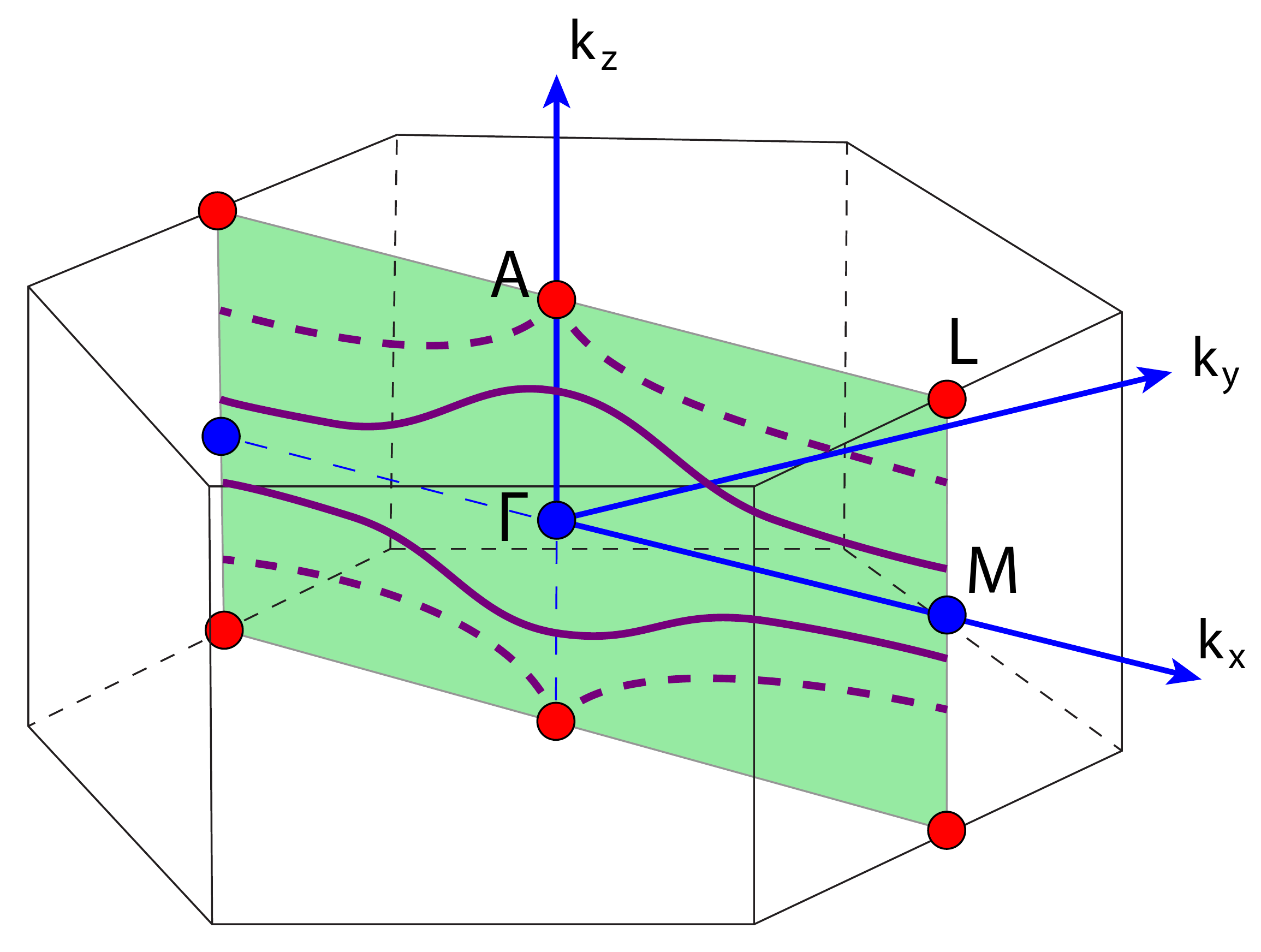} 
\vspace{0.8cm} }
\subfigure[]{
\includegraphics[width = 0.432\columnwidth]{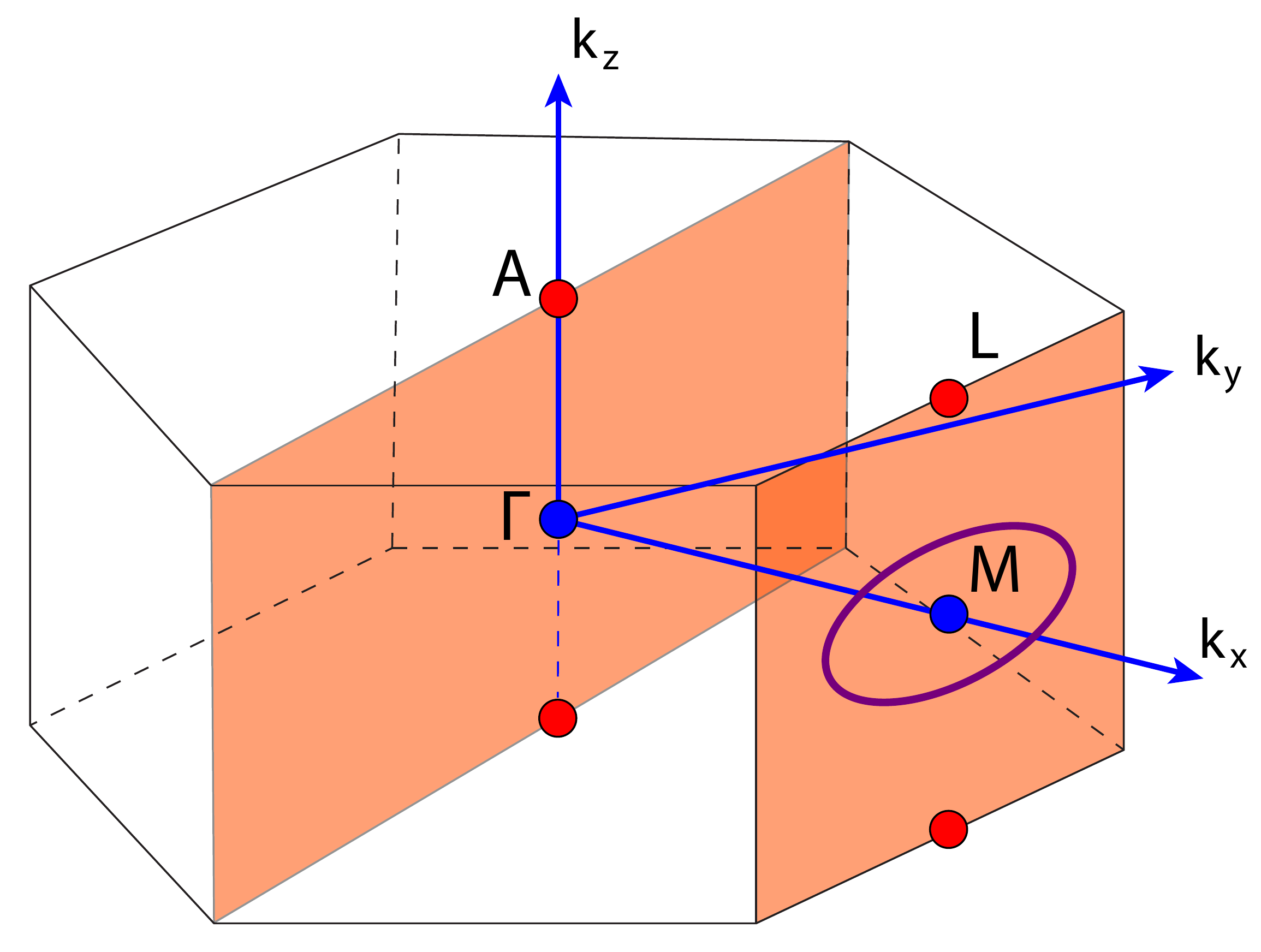}
} 
\subfigure[]{
\includegraphics[width = 0.432\columnwidth]{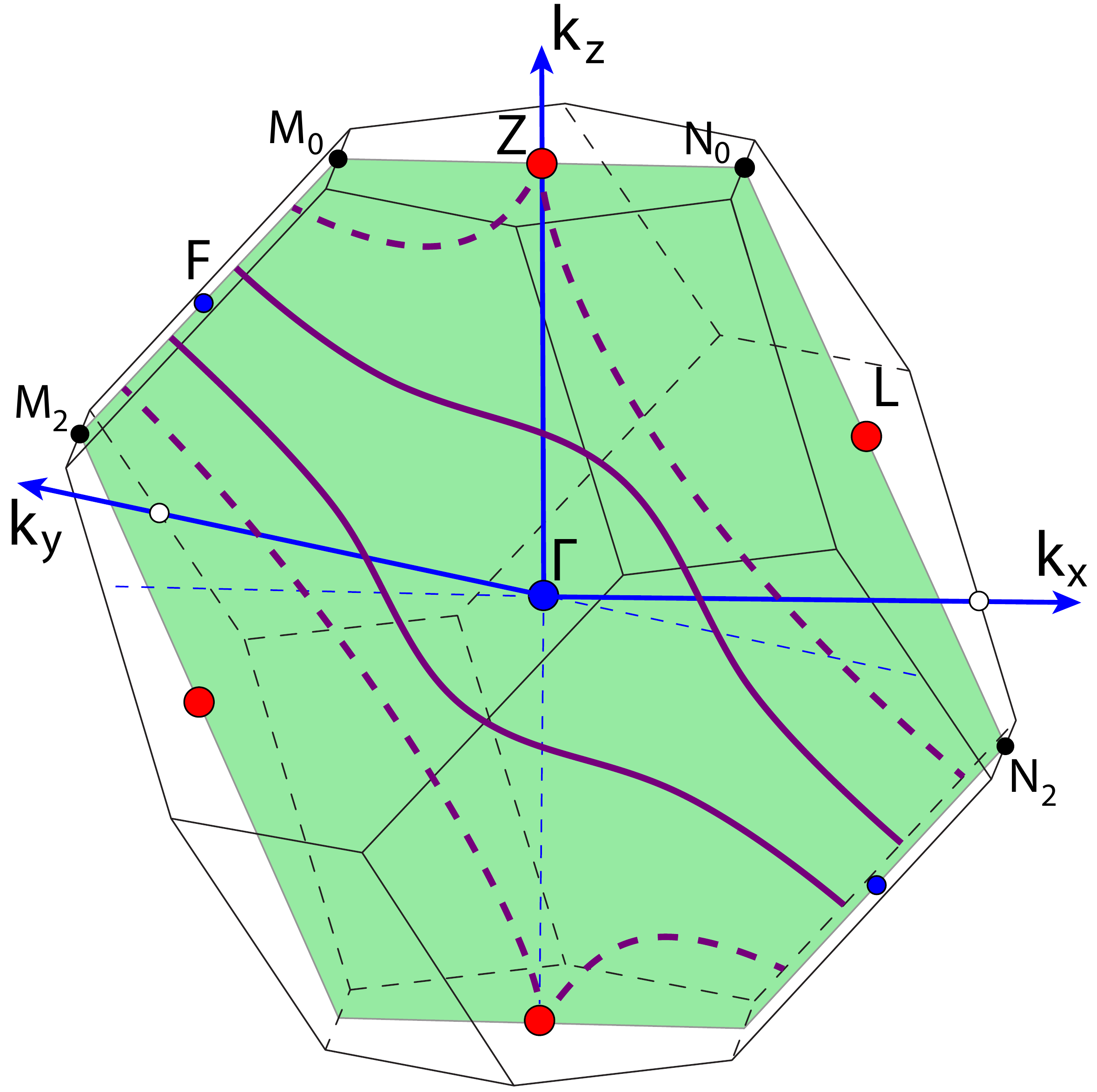} } 
\caption{ \label{fig:mir_bz}  \label{mFig2}
(a)
Band connectivity diagram for a path within the  $k_{x} k_{z}$-plane of the hexagonal BZ of SG No.~158 connecting $\Gamma$ to $A$ [see panel (b)]. 
The color indicates the glide-mirror eigenvalues~\eqref{eq:mir_eigen} of the Bloch bands. %
(b)-(d)
Weyl nodal lines (purple solid and dashed curves) protected by glide-mirror symmetries for (b)
SG No.~158, (c) SG No.~159  and (d) SG No.~161  are shown in the respective bulk BZs. 
The TRIMs are indicated by the blue and red dots and the mirror planes are shown in green and orange.
For SG Nos.~158 and 161 there are two nodal lines separating the blue from the red TRIMs. Depending
on the dimensionality of the irreps at the $A$ point ($Z$ point) these nodal lines either touch (dashed curves) or do not touch (solid curves) the $A$ point ($Z$ point).
In SG No.~159 there is a nodal line in the $k_x = \pi$ plane enclosing a single TRIM.
Note that there are different possibilites for the shape and connectivity of the nodal lines; here
we only show the simplest one. 
}
\end{figure}

\section{Non-symmorphic nodal lines}
\label{mSec3}

Using the techniques discussed in the previous section we will now prove that glide-reflection symmetries lead to 
symmetry-enforced nodal lines in materials with trigonal space groups. As before, this derivation is base on
 (i) the momentum-dependence of the symmetry eigenvalues
and (ii) the compatibility relations between irreps.

\subsection{Symmetry eigenvalues}

We consider glide mirror symmetries $M$
with a fractional shift of one half of the lattice constant along the $z$ axis. 
Applying these symmetries twice we obtain $M^2 = - \hat{z}$, where the minus sign is due to 
the $2\pi$ rotation of the electron spin. From this it follows that the eigenvalues
of the Bloch wavefunctions in the mirror planes are given by
\begin{equation}
\label{eq:mir_eigen}
M \ket{\psi_{\pm} (\kk)} = \pm i e^{-i k_{z}/2} \ket{\psi_{\pm} (\kk)} \, .
\end{equation}
The mirror planes contain either two or four TRIMs at which the Bloch
bands are Kramers degenerate (blue and red dots in Fig.~\ref{mFig2}).
It follows from Eq.~\eqref{eq:mir_eigen} that the $M$ eigenvalues of the Bloch bands
at the TRIMs $A$, $L$, and $Z$ are $\pm 1$, while at the TRIMs $\Gamma$, $M$, and $F$ they are $\pm i$.  
Due to the anti-unitary time-reversal symmetry,
at $\Gamma$, $M$, and $F$ the Kramers pairs are formed between bands with opposite $M$ eigenvalues (blue dots), 
while at $A$, $L$, and $Z$ they are formed between bands with the same eigenvalues (red dots).
Hence, as we move from a blue TRIM to a red TRIM the Kramers pairs must switch partners.
In the absence of additional symmetries this leads to a non-trivial band connectivity with a symmetry-enforced band crossing [see Fig.~\ref{mFig2}(a)].
Since this is true for any path connecting a blue TRIM to a red TRIM, the Weyl band crossings occur
along a one-dimensional line separating the blue TRIMs from the red TRIMs [see Figs.~\ref{mFig2}(b)-(d)]. 
We note that the presence of additional symmetries, in particular  inversion, leads to additional
degeneracies, thereby invalidating the above argument. Thus, symmetry-enforced nodal
lines only occur in SGs with a mirror-glide symmetry but no inversion symmetry.
Among the trigonal SGs there are only three SGs satisfying these criteria,
namely Nos. 158, 159, and 161.

\begin{table}[t!]
\begin{ruledtabular}
\begin{tabular}{ c c }  
Irreps/Sym. Element 
& 
$M_{010}$
\Tstrut\Bstrut\\ \hline
$\bar{\Gamma}_{4}$ & $-i$ \Tstrut\\
$\bar{\Gamma}_{5}$ & $+i$ \\
$\bar{\Gamma}_{6}$ & 
$\begin{pmatrix}
0 & e^{-i\pi/3} \\
e^{-i2\pi/3} & 0
\end{pmatrix}  $\\ \hline 
$\bar{A}_{4}$ & $+1$ \Tstrut\\
$\bar{A}_{5}$ & $-1$ \\
$\bar{A}_{6}$ & 
$
\begin{pmatrix}
0 & e^{i\pi/6} \\
e^{-i\pi/6} & 0
\end{pmatrix}
$ \\ \hline
$\bar{M}_{3}$ & $-i$ \Tstrut\\
$\bar{M}_{4}$ & $+i$  \\
\hline
$\bar{D}_{3}$ & $-i e^{ik_{z}/2}$ \Tstrut\\
$\bar{D}_{4}$ & $+i e^{ik_{z}/2}$  \\
\hline
$\bar{L}_{3}$ & $+1$ \Tstrut\\
$\bar{L}_{4}$ & $-1$ 
\end{tabular}
\end{ruledtabular}
\caption{
Double-valued irreps of SG No.~158 ($P3c1$) without time-reversal symmetry
at the TRIMs $\Gamma$, $A$, $M$, and $L$, and the plane $D$ (i.e., the $k_x k_z$-plane), see
Fig.~\ref{mFig2}(b).
We use the same convention as in Ref.~\cite{elcoro_aroyo_JAC_17} for labelling the irreps. 
\label{Tab158} \label{mTab4}
}
\end{table}

\begin{figure*}[th]
\centering
\includegraphics[width=\textwidth]{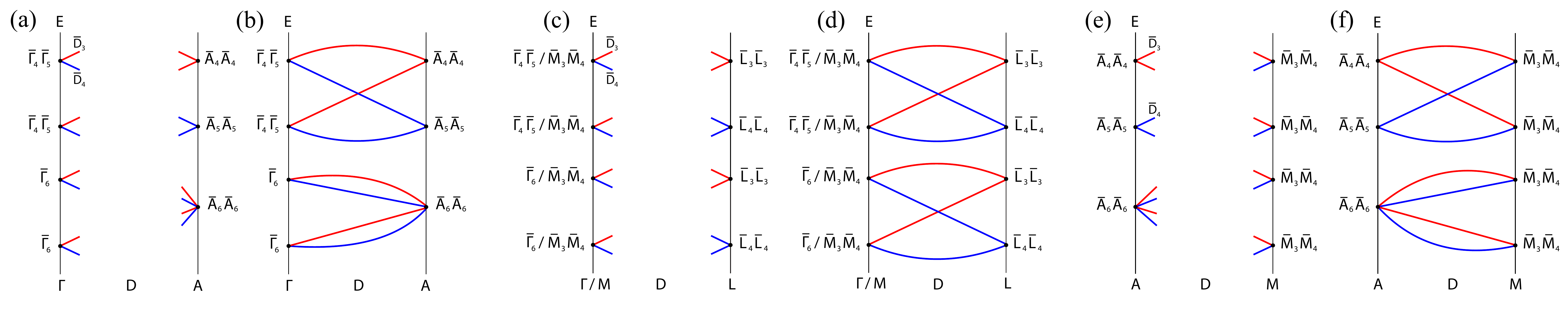}
\caption{ \label{fig:158irrep1}  \label{mFig3}
(a), (c), (e) Kramers pairings and compatibility relations for SG No.~158 ($P3c1$) 
between the irreps at the TRIMs and at a point in the mirror plane $D$ (i.e., a point in the $k_x k_z$-plane).
(b), (d), (f) Band connectivity diagrams for SG No.~158 for paths within the $D$ plane
connecting two TRIMs. Due to these nontrivial band connectivities
the $D$ plane must contain at least two Weyl nodal lines, see Fig.~\ref{mFig2}(b).
} 
\end{figure*}

SG No.~158 ($P3c1$) has  hexagonal lattice system (P-trigonal) and contains the glide-mirror symmetry
\begin{equation}
\label{eq_mir_SG_158}
M_{010} : (x,y,z) \rightarrow (x,x-y,z+ \frac{1}{2}) \otimes (- \frac{i\sqrt{3}}{2} \sigma_{x} - \frac{i}{2} \sigma_{y} ) \, ,
\end{equation}
which leaves the $k_{x} k_{z}$-plane invariant.
Following the above arguments we find that this glide-mirror symmetry leads to two nodal loops in the $k_{x} k_{z}$-plane,
which separate the red TRIMs from the blue TRIMs, as shown in Fig.~\ref{fig:mir_bz}(b). 
If the irrep at the A point has dimension four (instead of two), these nodal lines touch
the A point [dashed curve in Fig.~\ref{fig:mir_bz}(b)].
We note that  these nodal loops must be symmetric under
time-reversal symmetry and the point-group operations of SG No.~158, but
are otherwise free to move within the mirror plane.

SG No.~159 ($P31c$) has also hexagonal lattice system (P-trigonal) and contains the glide-mirror symmetry
\begin{equation}
\label{eq:m210}
M_{210} : (x,y,z) \rightarrow (-x, -x+y, z + \frac{1}{2}) \otimes (\frac{i}{2} \sigma_{x} + \frac{i\sqrt{3}}{2} \sigma_{y} ) \,,
\end{equation}
which leaves both the $k_x =0$ and the $k_x = \pi$ planes invariant, see Fig.~\ref{mFig2}(c). 
Both mirror planes have two TRIMs, between which the Kramers pairs change partners. Hence,
according to the above arguments, there should be nodal lines in both mirror planes, which encircle
either the $\Gamma$/$M$ or $A$/$L$ points. However,
at the $A$ point there exist four-dimensional irreps, leading to a four-fold band degeneracy.
For this reason, symmetry-enforced nodal lines generically occur only in the $k_x=\pi$ plane, 
but not in the $k_x=0$ plane
\footnote{We note that at the A point there exist both two-dimensional and four-dimensional irreps. Hence, nodal lines are also possible in the $k_x=0$ plane,
but they are not guaranteed to exist by symmetry alone.}.

SG No.~161 ($R3c$) has rhombohedral lattice system (R-trigonal)  and contains the glide mirror symmetry
\begin{equation}
\label{eq:m110}
M_{110} : (x,y,z) \rightarrow (-y, -x, z + \frac{1}{2}) \otimes (- \frac{i\sqrt{3}}{2} \sigma_{x} - \frac{i}{2} \sigma_{y} ) \,,
\end{equation}
which leaves the plane $k_{x}=-k_{y}$ invariant, see Fig.~\ref{mFig2}(d).
As in SG No.~158, this glide-mirror symmetry gives rise to two nodal lines separating
the two red TRIMs  from the two blue TRIMs. 
If the bands transform at the $Z$ point  under an irrep with dimension four (instead of two), then the nodal lines
touch the $Z$ point [dashed curve in Fig.~\ref{mFig2}(d)].

\subsection{Compatibility relations between irreps}

The existence of symmetry-enforced nodal lines can also be derived from the compatibility relations
between irreps. Here, we present this derivation for SG No.~158 ($P3c1$). The derivation for SG Nos.~159 and 161 
proceeds in a very similar way.

For SG No.~158 the relevant high-symmetry plane is the plane $D$ (i.e., the $k_x k_z$-plane), which is left invariant
under the glide-mirror symmetry $M_{010}$, Eq.~\eqref{eq_mir_SG_158}.  This mirror plane contains
the four TRIMs $\Gamma$, $A$, $L$, and $M$. To determine the connectivity of the bands 
we first need to derive the irreps at these four TRIMs and at the $D$ plane. The double-valued irreps
without time-reversal symmetry are listed in Table~\ref{mTab4}. We find
that these  irreps are all complex with the exception
of $\bar{\Gamma}_6$, which is pseudoreal.
Pseudoreal irreps are time-reversal invariant by themselves,
while complex irreps need to be paired up to construct
time-reversal invariant irreps~\cite{bradley_irreps,miller_lover_irreps} (cf.\ Fig.~\ref{mFig3}).
Note that at the $A$ point there are both two-dimensional and four-dimensional 
time-reversal invariant irreps. 
The decomposition of these time-reversal invariant irreps, as we move from a TRIM to a nearby point in the mirror plane~$D$,
can be inferred from Eq.~\eqref{eq:chara}, or alternatively from the program \texttt{DCOMPREL} 
on the Bilbao Crystallographic Server~\cite{elcoro_aroyo_JAC_17}. 
The resulting compatibility relations are depicted in Figs.~\ref{mFig3}(a), \ref{mFig3}(c), and \ref{mFig3}(e). 
Using these compatibility relations, we can now construct the band connectivity diagrams
for a path within the $D$ plane that connects two TRIMs, see Figs.~\ref{mFig3}(b), \ref{mFig3}(d), and \ref{mFig3}(f).
We observe that for any path connecting $\Gamma$/$M$  to $L$, there must be at least one band crossing.
The same is true for paths connecting $\Gamma$ to $A$ or $M$ to $A$,
albeit here the band crossings can be pinned at the $A$ point, if the corresponding irrep is four dimensional.
Hence, the $D$ plane must contain at least two Weyl nodal lines  enclosing the $\Gamma$ and $M$ points, 
as shown in Fig.~\ref{mFig2}(b).

 \section{Filling constraints}
\label{sec_filling_constraints}

According to band theory, a noninteracting band insulator can only exist
if the electron filling $\nu$ is an even integer, i.e.,  $\nu \in 2 \mathbbm{N}$. 
Vice versa, materials with $\nu \notin  2 \mathbbm{N}$ must necessarily by
(semi-)metals.
In materials with nonsymmorphic symmetries, however, these
filling constraints are more rigid. 
This is  because nonsymmorphic symmetries
enforce band crossings, that lead to groups
of more than two connected bands.
Thus, using the analysis form Secs.~\ref{mSec2} and~\ref{mSec3}
we can derive the more rigid filling constraints. 
For example, for SG No.~144 (P3$_1$)
we observe that along the $\Gamma - \Delta - A$ line
six bands form a connected group.
Thus, in an insulator with SG No.~144 this group
of six bands must be fully filled, i.e., the electron 
filling $\nu$ must be an element of $6 \mathbbm{N}$.
Using these arguments we have derived
the filling constraints for all the nonsymmorphic trigonal 
space groups,  see fourth column of Table~\ref{mTab1}.
These results are in full agreement with the analysis of Ref.~\cite{watanabe_vishwanath_PRL_16}. 


\section{Example Materials}
\label{mSec4}

Having identified the SGs whose nonsymmorphic symmetries
enforce Weyl points and Weyl nodal lines, 
we can now perform a database search for
materials in these SGs.
We look for suitable compounds in the ICSD database~\cite{ICSD_link}, the AFLOW database~\cite{aflow_citation}, and the Materials Project
database~\cite{materials_project_citation,materials_project_web_link},
which yields four materials with accordion Weyl points and one material with Weyl nodal lines. 

For each of these example materials we perform electronic band structure calculations
with the Vienna \textit{ab initio} simulation package (VASP) \cite{Kresse1996,Kresse1996-2}, using
the projector augmented wave (PAW) method~\cite{Bloechl1994,Kresse1999} and the 
PBE~\cite{PBE_PRL_96} and mBJ~\cite{mBJ_JCP_06,blaha_mBJ_PRL_09} exchange-correlation functionals.
As input for the \textit{ab initio} calculations we use the experimental crystal structures of 
Refs.~\cite{teske_Cu2SrSnS4,teske_Cu2SrGeS4,tordjman_Ag2HPO4_1978,adenis_tellurium_acta_crystallographica_89,jaussaud_Te16Si38_solid_state_science_04}.
In the following we discuss the band structures along those high-symmetry lines/planes where
the predicted band crossings occur. The full band structures along all high-symmetry directions of the BZ
are presented in Appendix~\ref{appencix_band_structure}.
For two examples we also compute surface states using tight-binding models
constructed from Wannier functions~\cite{mostofi_wannier_08}.

\subsection{Materials with Weyl nodal points}
\label{sec_materials_Weyl_points}

Here, we present four materials which exhibit Weyl nodal points with   accordion-like
dispersions.

 \begin{figure}[t!]
\includegraphics[width = 1.0\columnwidth]{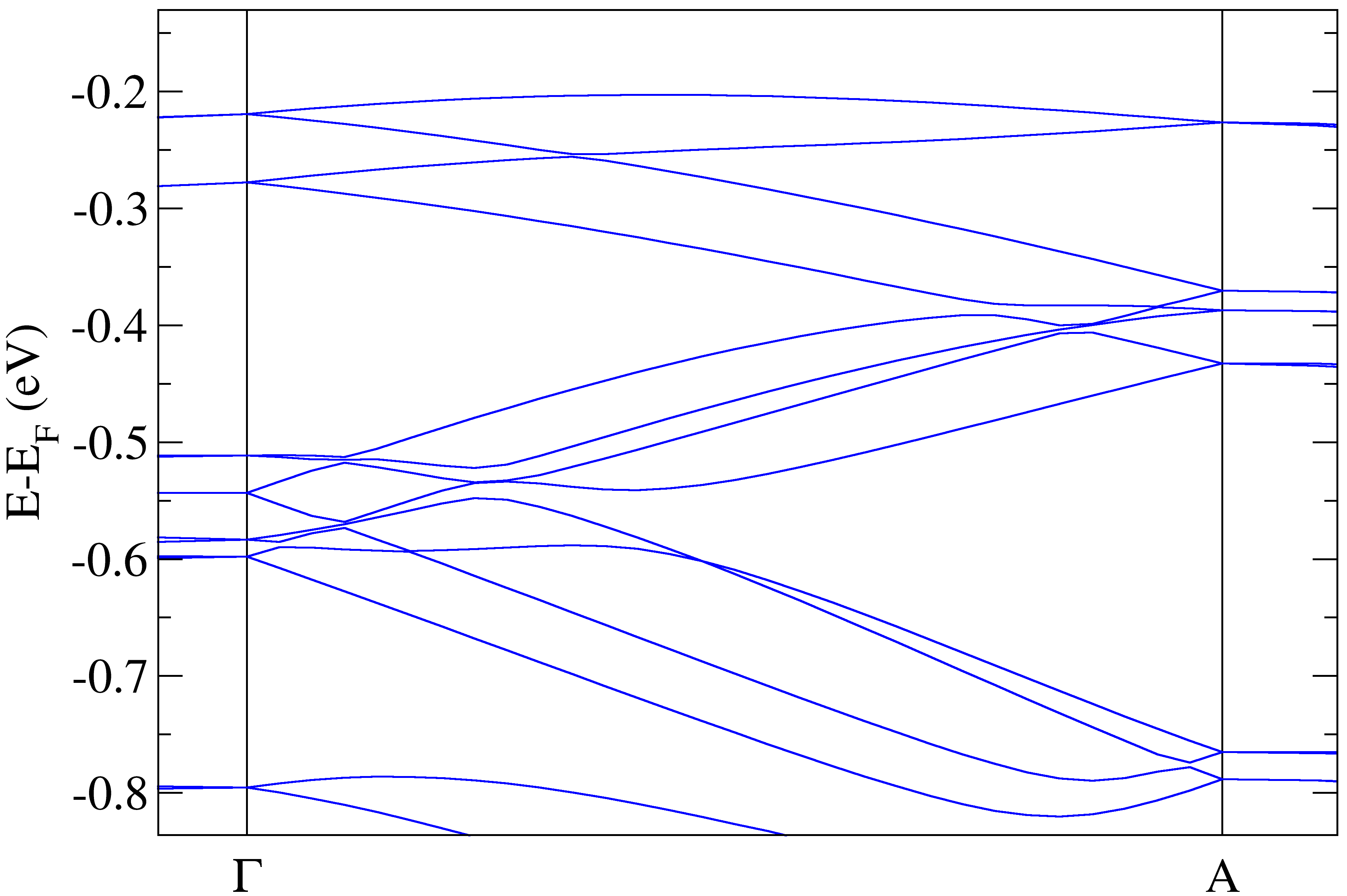}
\caption{
\label{fig_Cu2SrSnS4}
\label{mFig4}
Electronic band structure of Cu$_2$SrSnS$_4$ in SG No.~144 ($P3_1$)
computed with the mBJ functional~\cite{mBJ_JCP_06,blaha_mBJ_PRL_09}. 
The band crossings along the  $\Gamma$--$\Delta$--A   line are symmetry-enforced
by the screw rotation $C_{3,1}$, see Sec.~\ref{mSec2}.  
}
\end{figure}

\paragraph{Cu$_2$SrSnS$_4$ and Cu$_2$SrGeS$_4$---} 
The  chalcogenide Cu$_2$SrSnS$_4$, which crystallizes in SG No.~144 ($P3_1$)~\cite{teske_Cu2SrSnS4,tong_Cu2SrGeS4_elsevier},
is an example of a material with Weyl points along the  $\Gamma$--$\Delta$--$A$ line, 
as discussed in Sec.~\ref{mSec2}. 
The electronic band structure computed with the mBJ functional, see Figs.~\ref{mFig4} and \ref{mFigA1}(a), shows
a band gap of about 1.1~eV, which is somewhat smaller than the experimentally reported value 
of 1.78~eV~\cite{tong_Cu2SrGeS4_elsevier}. 
We observe in Fig.~\ref{mFig4} that there are 
 groups of $6n$ connected bands between  $\Gamma$ and $A$,
which form a large number of Weyl points.  
This is in full agreement with the band connectivity diagrams of Figs.~\ref{mFig1}(b)
and~\ref{mFig1}(c). 
The Weyl points of Cu$_2$SrSnS$_4$ are 
symmetry protected by the screw rotation $C_{3,1}$, Eq.~\eqref{eq:c3}. 
Their topological stability is ensured by 
quantized Chern numbers. Due to the bulk boundary correspondence,
these   Chern numbers lead to arc surface states.
Unfortunately, due to the large band gap 
of $\sim 1.8$~eV, it will be very challenging to
probe  the bulk Weyl points or arc surface states
of Cu$_2$SrSnS$_4$ using photoemission, tunneling spectroscopy,
or transport experiments. Perhaps, one could
try to grow  Cu$_2$SrSnS$_4$ as a thin film on a conducting
substrate, which might allow to probe the Weyl points using photoemission.

Cu$_2$SrGeS$_4$ is very similar to Cu$_2$SrSnS$_4$. It crystallizes in SG No.~145 ($P3_2$)~\cite{teske_Cu2SrGeS4}
and its band structure shows a large number of Weyl points
along the $\Gamma$--$\Delta$--$A$ line, see Fig.~\ref{mFigA1}(b) in Appendix~\ref{appendixA}.

\begin{figure}[t!]
\includegraphics[width = 1.0\columnwidth]{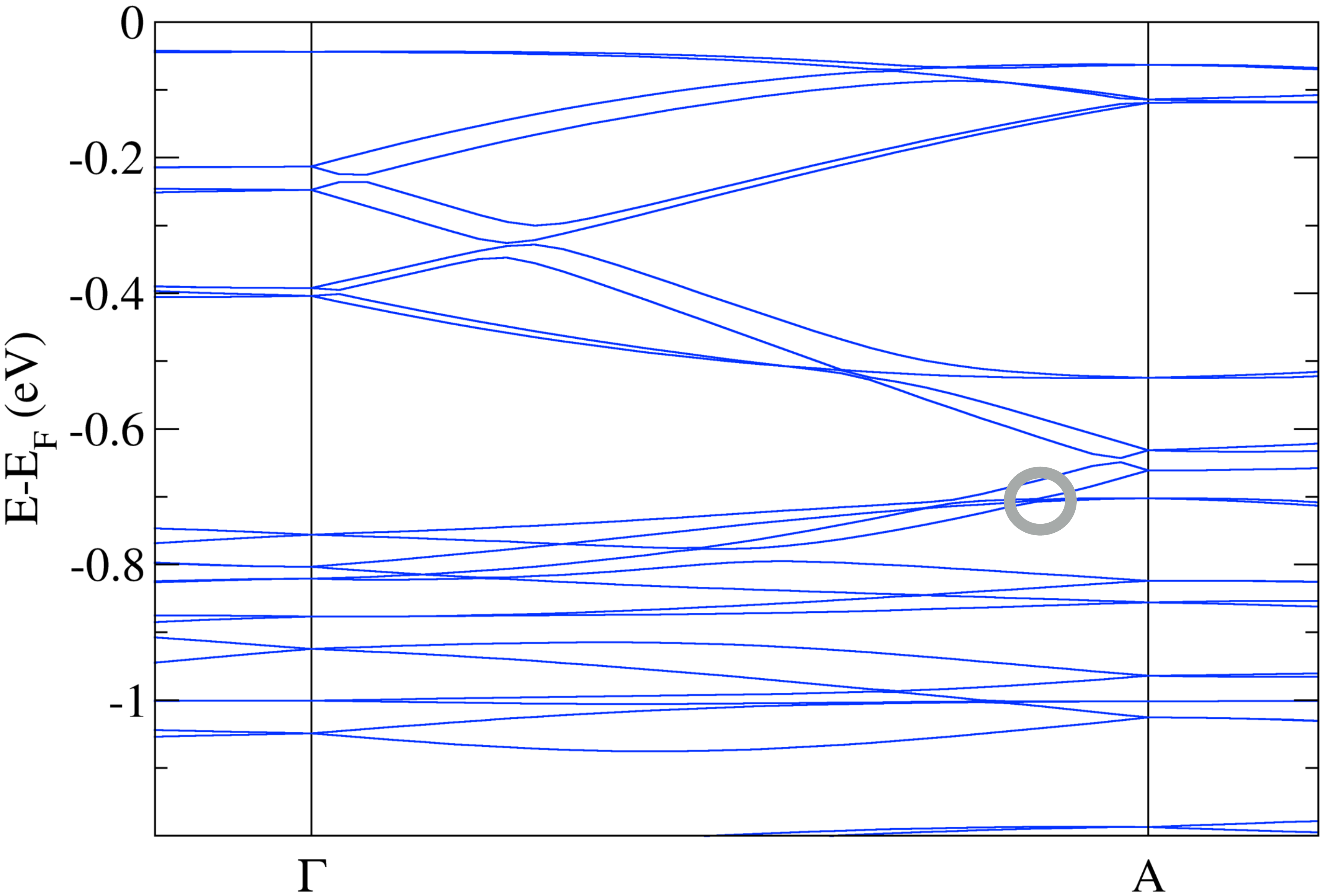}
\caption{
\label{fig_Ag2PHO4}
\label{mFig5}
(a) First-principles band structure of silver hydrogen phosphate Ag$_2$HPO$_4$
in SG No.~151 ($P3_112$) computed with the mBJ functional. The band crossings
along the  $\Gamma$--$\Delta$--$A$ line are Weyl points
that are symmetry enforced by the screw rotation $C_{3,1}$ (see Sec.~\ref{mSec2}). 
}
\end{figure}

\begin{figure}[t!] 
\includegraphics[width = 1.0\columnwidth]{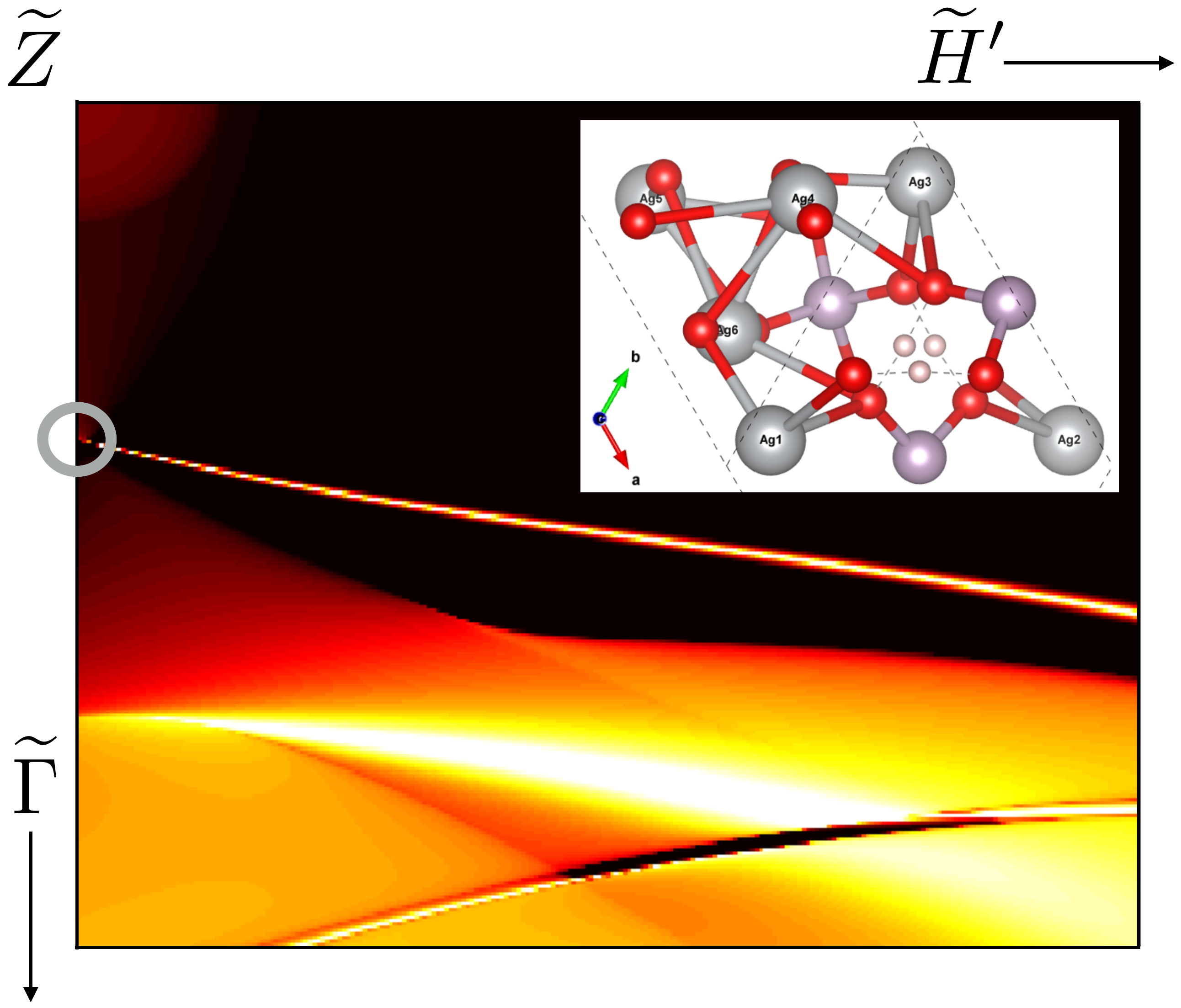}
\caption{
\label{fig_Ag2PHO4_SS}
Momentum-dependent surface density of states
for the (-110) surface of Ag$_2$HPO$_4$  
at the energy $E-E_F=-0.704$~eV.
The crystal is terminated such that
the  silver atoms ``Ag3'', ``Ag4'', and ``Ag5"  are at the top layer (see crystal structure  in inset).
Yellow and black correspond to high and low densities, respectively.
The location of the projected Weyl point  is marked by a gray circle,
cf.~Fig.~\ref{mFig5}.
For a plot of the (-110) surface BZ see Fig.~\ref{mFig1}(a). 
}
\end{figure}

\paragraph{Ag$_2$HPO$_4$---}
Silver hydrogen phosphate with the chemical formula Ag$_2$HPO$_4$ is a salt, that crystallizes in SG No.~151~\cite{tordjman_Ag2HPO4_1978}.
According to our analysis of Sec.~\ref{mSec2},   band structures with this SG symmetry exhibit Weyl points  
along the $\Gamma$--$\Delta$--$A$ line.
Figures~\ref{fig_Ag2PHO4} and~\ref{mFigA1}(c) display the first-principles
band structure of Ag$_2$HPO$_4$, computed with the mBJ functional.
The band structure shows a band gap of about $2.8$~eV. 
Along the  $\Gamma$--$\Delta$--$A$ direction
we observe a group of 24 ($= 6 \times 4$) connected bands,
with many crossings, similar to the band connectivity diagrams of Fig.~\ref{mFig1}.
These band crossings are Weyl points, whose topology is characterized by a nonzero Chern  number. 
The nontrivial topology of these Weyl points gives rise to arc states at the surface, which
connect Weyl points with opposite chirality. 
In Fig.~\ref{fig_Ag2PHO4_SS}
we demonstrate this
for  the Weyl point marked by the gray circle in Fig.~\ref{mFig5}.
At the (-110) surface we observe an arc state that
emanates out from the projected Weyl point.

While the \emph{ab-initio} band structure of Ag$_2$HPO$_4$ nicely illustrates the predicted accordion states, 
this material is rather unsuitable for experimental investigations. Its large band gap and the
fact that it cannot be doped easily makes it experimentally impossible to probe the bulk Weyl points
and arc surface states.

\paragraph{Te---} \label{paragraph_tellurium}
At ambient conditions, elemental tellurium (Te) has a trigonal crystal structure with SG No.~152 ($P3_121$) ~\cite{tellurium_cherin_acta_crystallographica_67,adenis_tellurium_acta_crystallographica_89}.
In this structure the Te atoms form helical chains along the $z$ direction, which are arranged in a hexagonal array
\footnote{Note that depending on the chirality of these chains Te can also be in SG No.~153 ($P3_212$).}.
Recent first principles calculation have shown that the valence and conduction bands of Te at the $H$ point exhibit
several Weyl points~\cite{tellurium_Weyl_hirayama_PRL_15,tellurium_ARPES_17}. Moreover,
it was found that hydrostatic or uniaxial strain leads to a band inversion
at the $H$ point, thereby transforming Te  into a strong topological insulator~\cite{tellurium_TI_Ong_PRL_13}.
 
\begin{figure}[t!]
\includegraphics[width = 1.0\columnwidth]{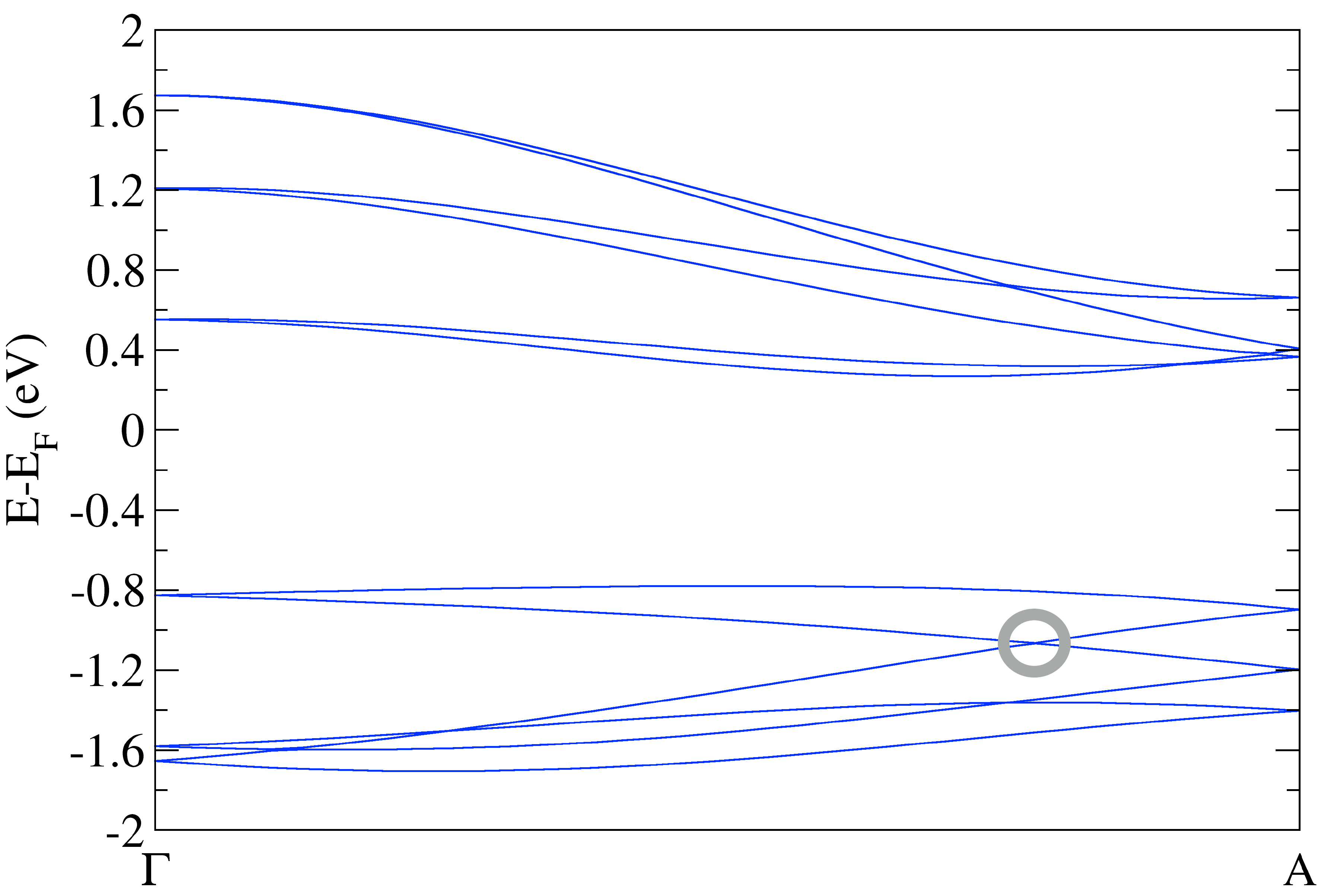}
\caption{
\label{fig_band_struct_Te}
\label{mFig6}
Electronic band structure   
of trigonal tellurium in SG No.~152 ($P3_121$), computed with the PBE functional. 
The band crossings with accordion dispersions 
along the $\Gamma$--$\Delta$--A line
are symmetry enforced by the screw rotation $C_{3,1}$ (see Sec.~\ref{mSec2}).
}
\end{figure}

It follows from our analysis of Sec.~\ref{mSec2}, that Te exhibits Weyl points not only at the $H$ point, but also
along the $\Gamma$--$\Delta$--$A$ line, which is left invariant by the screw rotation $C_{3,1}$. 
This is clearly visible in the \emph{ab-initio} band structure calculations of Figs.~\ref{mFig6} and~\ref{mFigA1}(d), which show
Weyl points with accordion-like dispersions between $\Gamma$ and $A$. As before, we observe that
the band connectivity fully agrees with the theoretical prediction of Fig.~\ref{mFig1}. 
By the bulk boundary correspondence, the Weyl points   at the $H$ point and on the $\Gamma$--$\Delta$--$A$ line lead
to numerous arc surface states, see Appendix~\ref{appendixB}. 
These arc surface states should be readily observable in photoemission and scanning tunneling experiments.

In closing, we note that trigonal selenium (Se), which also crystalizes in SG No. 152 ($P3_121$)~\cite{cherin_trigonal_slenium}, has a very similar band structure
to tellurium, albeit the spin-orbit coupling is considerably smaller [Fig.~\ref{mFigA1}(e)]. Therefore, the bands forming
Weyl points along the $\Gamma$--$\Delta$--A line of Se are only separated by  less than $\sim 50$ meV. This makes the experimental detection of 
the accordion Weyl  points in Se more  challenging than in Te.   
 
\subsection{Weyl nodal lines in Te$_{16}$Si$_{38}$}

\begin{figure}[thp!]
\includegraphics[width = 1.0\columnwidth]{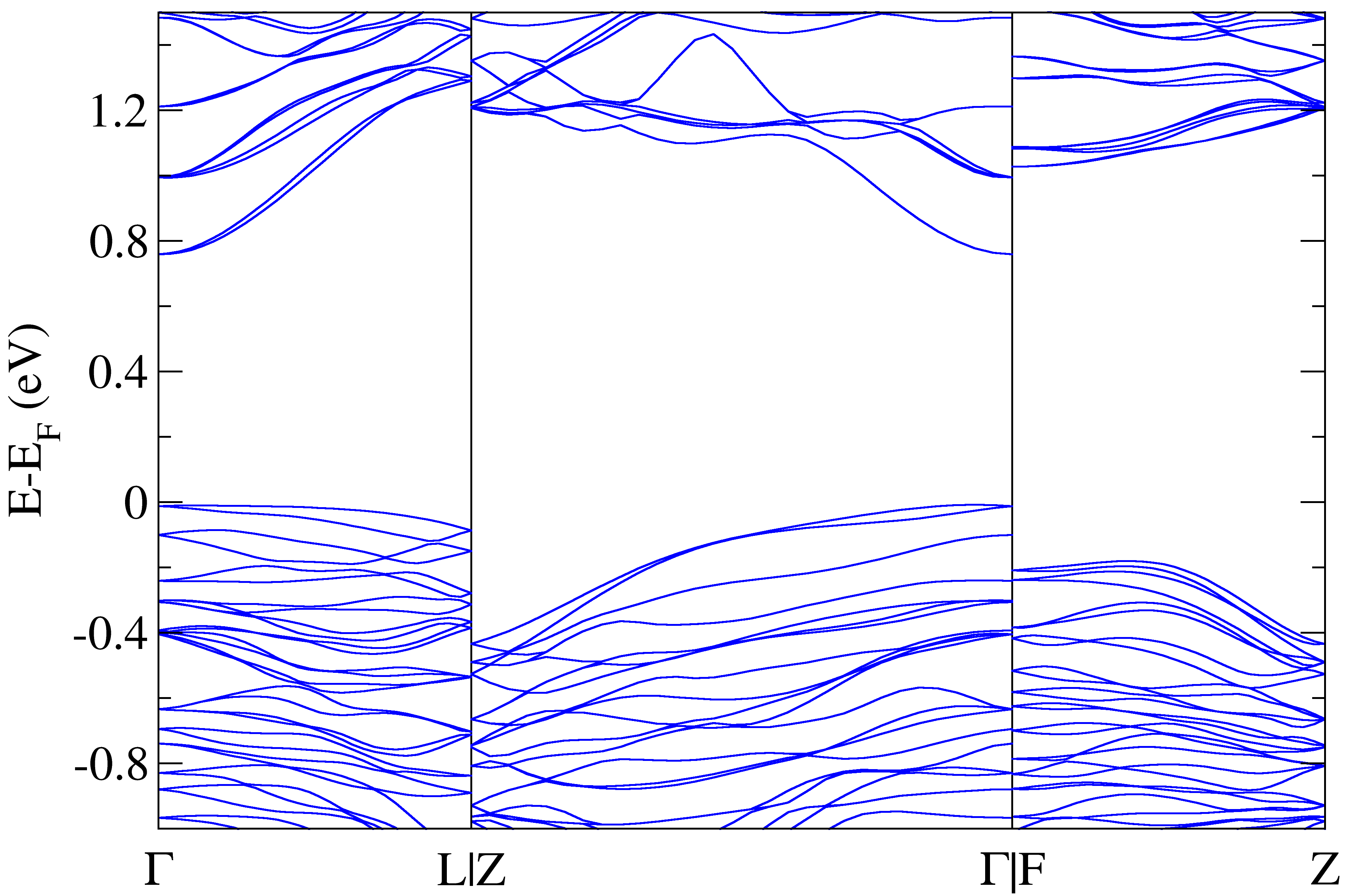}
\caption{ \label{fig_zoom_Te16Si38}
DFT-PBE band structure of rhombohedral Te$_{16}$Si$_{38}$
along high-symmetry paths of the $k_x = - k_y$ plane of
the rhombohedral BZ  [see Fig.~\ref{mFig2}(d)].
The  band crossings are part of Weyl nodal lines, which are protected
by the glide mirror symmetry $M_{110}$.
}
\end{figure}

The tellurium-silicon clathrate Te$_{16}$Si$_{38}$ crystallizes in two different forms. 
While the low-temperature form is cubic, the high-temperature form is rhombohedral
with trigonal SG No.~161 $(R3c)$~\cite{jaussaud_Te16Si38_solid_state_science_04,jaussaud_inorganic_chemistry_05}.
According to our analysis of Sec.~\ref{mSec3}, materials in this SG exhibit Weyl nodal lines 
that are protected by the glide mirror symmetry $M_{110}$. 
In Fig.~\ref{fig_zoom_Te16Si38} we present the first principles band structure (with PBE functional)
of  rhombohedral Te$_{16}$Si$_{38}$ along high-symmetry paths
 of the $k_x = - k_y$ plane, which is left invariant by $M_{110}$ [see also Fig.~\ref{mFigA1}(f)].
We observe that along any path that connects  a blue TRIM to a red TRIM in the BZ of Fig.~\ref{mFig2}(d),
$4n$ bands form a connected group with many Weyl crossings.
These Weyl crossings are part of   Weyl nodal lines that separate 
the blue from the red TRIMs in the $k_x = - k_y$ plane.
The stability of these Weyl nodal lines is guaranteed
by a $\pi$-Berry phase, which
by the bulk-boundary correspondence gives
rise to drumhead surface states~\cite{chiu_schnyder_arXiv_2018}.
In addition, the bands of rhombohedral Te$_{16}$Si$_{38}$ 
carry a nonzero Berry curvature which becomes especially large
close to the Weyl nodal lines.

\section{Conclusions}
\label{sec_conclusion}
 
In summary, we have performed  a systematic analysis of 
nonsymmorphic band crossings in
nonmagnetic trigonal materials with strong spin-orbit coupling. 
We have found that among the 25 trigonal space groups, 
six support Weyl points, while three support Weyl nodal lines (see Table~\ref{mTab1}).
The electronic bands at these Weyl points
and Weyl lines are required to cross due to 
nonsymmorphic mirror or rotation symmetries alone, regardless
of the chemical composition and other material details.  
Hence, these Weyl crossings
occur in all bands of any material crystallizing in the space
groups listed in Table~\ref{mTab1}. 
Using this insight, we have performed a database search
and identified several existing materials that 
possess the predicted topological band crossings,
see last column in Table~\ref{mTab1}. 
The nontrivial topology of the band crossings
in these materials leads to a number 
of interesting and experimentally observable phenomena,
e.g., arc and drumhead surface states,
anomalous magneto-transport properties, and
anomalous Hall effects.
Particularly interesting are the nonsymmorphic band crossings and surface states of tellurium
(Figs.~\ref{fig_band_struct_Te} and \ref{mFigB1}).
We hope that our findings will encourage
experimentalists to study the topological 
properties of these compounds.

\acknowledgments
The authors thank C.~Ast, K.~von Klitzing, A.~Topp,  M.~G.~Vergniory, and A.~Yaresko
for useful discussions. 
B.K. thanks the Max Planck Institute for Solid State Research in Stuttgart for financial support.
C.-K.C. is supported by the Strategic Priority Research Program of the Chinese Academy of Sciences (Grant XDB28000000).
Work at Princeton was supported by NSF through the Princeton Center for Complex Materials, a Materials Research Science and Engineering Center DMR-1420541.


 \vspace{1.0cm}

 \appendix

\section{Additional band structure calculations}
\label{appencix_band_structure} 
\label{appendixA}

 \begin{figure*}[thp!]
\includegraphics[width = 2.0\columnwidth]{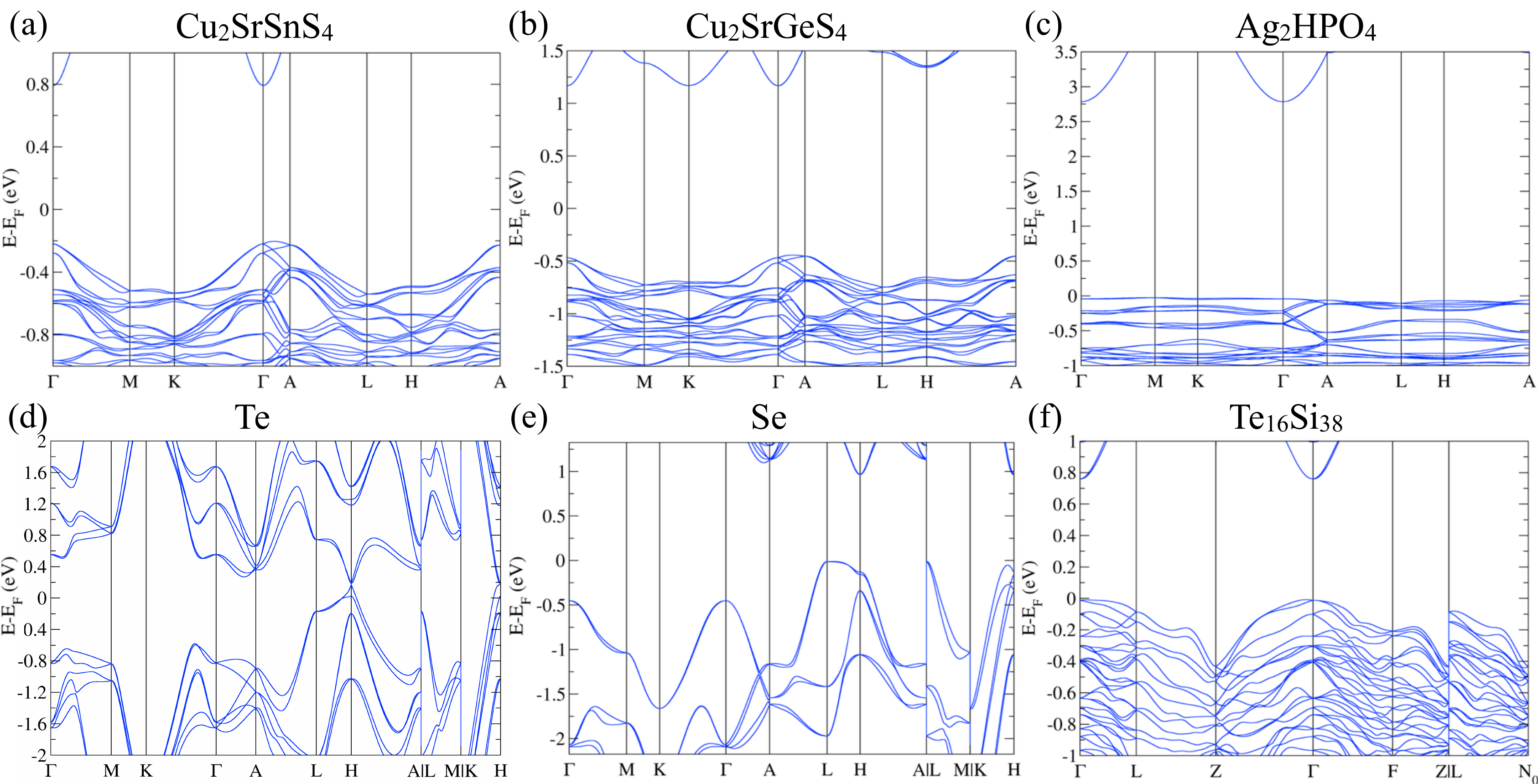}
\caption{ \label{fig_full_band_structures} \label{mFigA1}
(a), (b), (c) DFT band structures computed with the mBJ functional for Cu$_2$SrSnS$_4$, Cu$_2$SrGeS$_4$, and Ag$_2$HPO$_4$.
Along the $\Gamma-\Delta-A$ line, these materials exhibit Weyl points with accordion dispersions. 
For a zoom in of the band structure near the Weyl points see Sec.~\ref{sec_materials_Weyl_points} in the main text.
(d), (e)  DFT-PBE band structure of trigonal tellurium and selenium. Along the $\Gamma - \Delta - A$ direction 
there are groups of six connected bands, which cross at several Weyl points.
(f) DFT-PBE band structure  for Te$_{16}$Si$_{38}$, which exibits
Weyl nodal lines within the $k_x = - k_y$ plane. For a zoom in of the band structure around the Weyl nodal lines see
Fig.~\ref{fig_zoom_Te16Si38} in the main text.
}
\end{figure*}

In this Appendix, we present additional band-structure calculations
for the example materials studied in Sec.~\ref{mSec4} of the main text.
In Figs.~\ref{mFigA1}(a), \ref{mFigA1}(b), and \ref{mFigA1}(c), we present the full
band structures of the materials   Cu$_2$SrSnS$_4$, Cu$_2$SrGeS$_4$, and 
Ag$_2$HPO$_4$, respectively,
which exhibit Weyl points along $\Gamma - \Delta - A$. 
These Weyl points are symmetry enforced by the three-fold $C_{3,p}$ 
screw rotations. In the chalcogenides 
Cu$_2$SrSnS$_4$ and Cu$_2$SrGeS$_4$
the band gap is about 1.5~eV, while in the salt Ag$_2$HPO$_4$ it is about 2.8~eV. This makes it challenging to observe the
Weyl points in photoemission or electron tunneling experiments.
However, it might be perhaps possible to grow thin films of Cu$_2$SrSnS$_4$ and Cu$_2$SrGeS$_4$
on a conducting substrate to introduce charge carriers.

Figures~\ref{mFigA1}(d) and \ref{mFigA1}(e) display the full band structures
of trigonal tellurium and selenium.  Along the $\Gamma - \Delta - A$ line there are groups
of six connected bands with accordion dispersions. 
Since tellurium  is metallic, it should be possible to observe its accordion dispersion
and Weyl points using angle-resolved photoemission experiments.
Tellurium has many Weyl points also away from the $\Gamma - \Delta - A$ line, 
for example, at the $H$ point~\cite{tellurium_Weyl_hirayama_PRL_15,tellurium_ARPES_17}. 
These give rise to numerous arc surface states, see Appendix~\ref{appendixB}.

Figure~\ref{mFigA1}(f) shows the full band structure of the
tellurium-silicon clathrate
 Te$_{16}$Si$_{38}$,
which exhibits Weyl nodal lines in the $k_x = - k_y$ plane.
We observe that along the paths $\Gamma - L$, $\Gamma - Z$, and $F - Z$, 
the bands form groups of $4 n$ connected bands with Weyl crossings. These Weyl crossings
are part of nodal lines that separate $\Gamma$ and $F$ from $Z$ and $L$ in 
the $k_x = - k_y$ plane of the rhombohedral BZ, see Fig.~\ref{mFig3}(d).


\section{Surface states of tellurium}
\label{appendixB}

\begin{figure}[thp]
\includegraphics[width = 1.0\columnwidth]{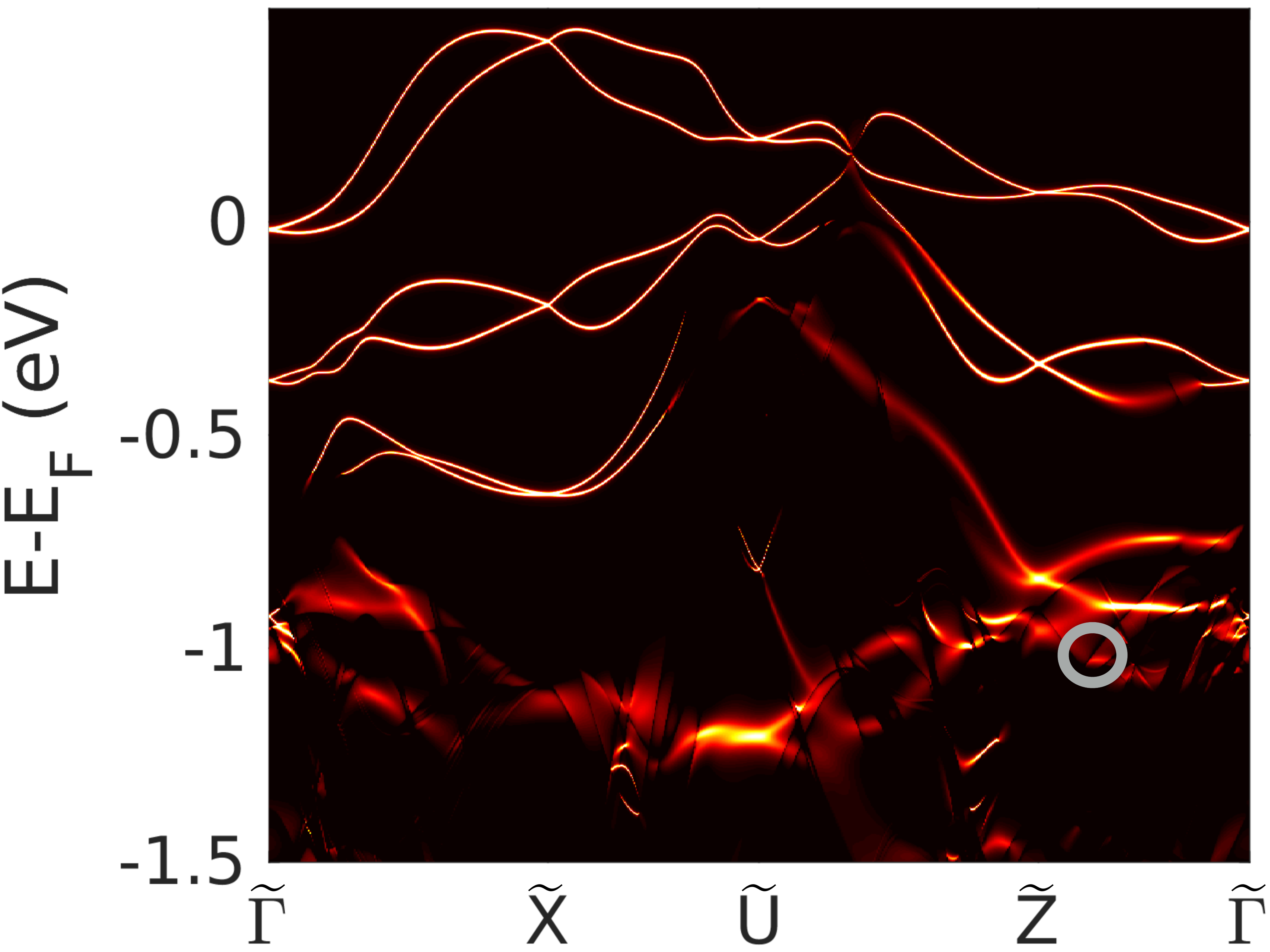}
\caption{ \label{fig_surface_state_tellurium} \label{mFigB1}
Energy dispersion of the surface states of trigonal tellurium for the (-110) surface
along high-symmetry momentum directions, cf.~Fig.~\ref{mFig1}(a).
The color grading indicates the surface density of states, with  yellow and  black
corresponding to high and low densities, respectively. 
The gray circle denotes the projected position of the Weyl point at the energy $E-E_F = -1.072$~eV, see Fig.~\ref{fig_band_struct_Te}. 
}
\end{figure}

Trigonal tellurium has many Weyl points and associated arc surface states.
In Fig.~\ref{mFigB1} we present the
surface state dispersion for the (-110) surface of Te~\footnote{Note that there is only one type of termination}.
The labels of the high-symmetry points in the (-110) surface BZ are given in Fig.~\ref{mFig1}(a).
We observe that there are several surface states emanating from the Weyl points near
the $H$ point. Remarkably, these surface states cross the entire surface BZ and lie within the bulk band gap,
in between the valence and conduction bands.
Since their energy is close to the Fermi level, 
they could play an important role for the \mbox{(re-)inter}\-pretation
of the unusual surface transport and photoemission measurements reported in Refs.~\cite{VONKLITZING19712201,photoemission_Te_Fitton_PRL_72,englert_klitzing_physica_status_solidi_77}. 

There are also arc surface states that are associated with the accordion bands discussed in Sec.~\ref{paragraph_tellurium} and Fig.~\ref{fig_band_struct_Te}.
For example, for the topmost Weyl point in the valence bands  at the energy $E-E_F = -1.072$ (gray circle in Fig.~\ref{fig_band_struct_Te}),
we observe an arc state that disperses towards the $\tilde{\Gamma}$ point.

\bibliographystyle{apsrev4-1}
\bibliography{trigonals_literature}

\appendix

\end{document}